\documentclass{aa} \usepackage{epsfig} \usepackage{amssymb}
\usepackage{times} \usepackage{aabib99} 

\newcommand{\n}{\ensuremath{\mem{n}}}
\newcommand{\p}{\ensuremath{\mem{p}}}
\newcommand{\ndr}{\ensuremath{^{13}\mem{N}}}
\newcommand{\hevi}{\ensuremath{^{4}\mem{He}}}
\newcommand{\czw}{\ensuremath{^{12}\mem{C}}}
\newcommand{\cdr}{\ensuremath{^{13}\mem{C}}}
\newcommand{\nvi}{\ensuremath{^{14}\mem{N}}}
\newcommand{\ose}{\ensuremath{^{16}\mem{O}}}
\newcommand{\osi}{\ensuremath{^{17}\mem{O}}}
\newcommand{\oac}{\ensuremath{^{18}\mem{O}}}
\newcommand{\fac}{\ensuremath{^{18}\mem{F}}}
\newcommand{\nezw}{\ensuremath{^{22}\mem{Ne}}}
\newcommand{\mgvi}{\ensuremath{^{24}\mem{Mg}}}
\newcommand{\mgfu}{\ensuremath{^{25}\mem{Mg}}}
\newcommand{\mgse}{\ensuremath{^{26}\mem{Mg}}}
\newcommand{\spr}{\mbox{$s$-process}}
\newcommand{\jahre}{\ensuremath{\, \mathrm{yr}}}

\newcommand{\beq}{\begin{equation}} \newcommand{\eeq}{\end{equation}}
\newcommand{\lsun}{\ensuremath{\, {\rm L}_\odot}}
\newcommand{\lhe}{\ensuremath{L_{\rm He}}}

\newcommand{\kelv}{\ensuremath{\,\rm K}}
\newcommand{\msun}{\ensuremath{\, {\rm M}_\odot}}
\newcommand{\mh}{\ensuremath{M_{\rm H}}} \newcommand{\etal}{et~al.\,}
\newcommand{\etalk}{et~al.,\,}
\newcommand{\apleq}{\ensuremath{\stackrel{<}{_\sim}}}

\newcommand{\mem}[1]{\ensuremath{\mathrm{ #1}}}
\newcommand{\dup}{dredge-up} \newcommand{\Dup}{Dredge-up}

\newcommand{\gradad}{\ensuremath{\nabla_\mathrm{ad}}}
\newcommand{\gradrad}{\ensuremath{\nabla_\mathrm{rad}}}
\newcommand{\nat}[2]{\ensuremath{#1 \cdot 10^{#2}}}

\newcommand{\kap}[1]{Sect.\,\ref{#1}}
\newcommand{\tab}[1]{Table\,\ref{#1}}
\newcommand{\gl}[1]{Eq.\,(\ref{#1})}
\newcommand{\abb}[1]{Fig.\,\ref{#1}}

\newcommand{\mdot}{\ensuremath{\dot{M}}}
\newcommand{\pdcz}{pulse-driven convective zone} \newcommand{\ms}{main
  sequence}

   \thesaurus{08         
              (08.16.4;  
               08.05.3;  
               08.09.3;  
               08.01.1)}
             \title{The evolution of AGB stars with convective
               overshoot}
             \titlerunning{AGB evolution with overshoot}
 
             \author{Falk Herwig \inst{1,2} } \authorrunning{F.
               Herwig}
             
             \institute{ Universit\"at Potsdam, Institut f\"ur Physik,
               Astrophysik, Am Neuen Palais 10,
               D-14469 Potsdam\\
               email: fherwig@astro.physik.uni-potsdam.de \and
               Astrophysikalisches Institut Potsdam (AIP),
               An der Sternwarte 16, D-14482 Potsdam\\
               }
             
             \date{Received Dec 1999; accepted June 2000}
\begin{document}
\maketitle

   \begin{abstract}
     The influence of extended convective mixing (overshoot) on
     asymptotic giant branch stellar evolution is investigated in
     detail.  The extended mixing is treated time-dependently, and the
     efficiency declines exponentially with the geometric distance
     from the convective boundary. It has been considered at all
     convective boundaries, including the He-flash convection zone in
     the intershell region which forms during the thermal pulses.
 
     Both the structural and the chemical evolution are affected by
     the inclusion of overshoot. The main results include a very
     efficient third dredge-up which leads to the formation of carbon
     stars of low mass and luminosity. A \cdr\ pocket which may serve
     as a neutron source for the \spr\ can form after the third
     dredge-up has reached into the \czw\ rich intershell.  Overshoot
     applied to the pulse-driven convective zone during the He-flash
     leads to a deeper penetration of the bottom of this convective
     zone into the C/O core below the He-burning shell. This in turn
     causes \hevi\ to be less abundant in the intershell while \czw\ 
     and \ose\ are more abundant compared to calculations without
     overshoot. We show that overshoot at the He-flash convection zone
     as well as at the base of the envelope convection enhance the
     efficiency of the third dredge-up. Characteristic properties of
     the structural and chemical evolution of AGB stars are presented.

     \keywords{Stars: AGB and post-AGB -- Stars: evolution -- Stars:
       interiors -- Stars: abundances }
       
   \end{abstract}

%

\section{Introduction}

Towards the end of their lifetime, stars of low and intermediate mass
($\mathrm{M}<8\msun$) evolve along the Asymptotic Giant Branch (AGB)
stage \cite{iben:83b,habing:96,lattanzio:97}.  The core of carbon and
oxygen is surrounded by a sandwich like structure consisting of a
helium burning shell, the hydrogen burning shell and the intershell
region in between. Evolved AGB stars undergo recurrent thermal
instabilities of the helium burning shell (He-flash) which are
referred to as thermal pulses (TP) \cite{schwarzschild:65,weigert:66}.
Locally, helium burning peak luminosities of $\log \lhe /\lsun \simeq
5 \dots 8$ are released and cause complex convective mixing events.
The He-flash causes a pulse-driven convective zone (PDCZ) in the
intershell region. After the He-flash the bottom boundary of the
convective envelope may engulf the deeper regions where material
previously synthesized by hydrogen and helium burning is present
(third dredge-up, TDUP).

Despite many studies of AGB evolution
\cite{iben:76,schoenberner:79,lattanzio:86,boothroyd:88c,vassiliades:93,bloecker:95a,d'antona:96,forestini:97,straniero:97,wagenhuber:98},
important details, like the surface enrichment with nuclear processed
material from the deep interior (dredge-up) or the origin of \cdr\ 
which is an important source of neutrons for the synthesis of heavy
elements in AGB stars, are not very well understood.  To improve this
situation we present the structural properties as well as the chemical
evolution of the interior of AGB stellar models with convective
overshoot.  Convective motions of matter approach the convective
boundary with a non-zero velocity and penetrate into the formally
stable region. These overshooting flows lead to extra mixing of
elements.  Canuto \cite*{canuto:98a} has pointed out that overshoot is
a dynamical consequence of Newton's laws and as such is unavoidable.
Previous studies have shown that models with overshoot can account for
several observed properties of AGB and post-AGB stars (Bl\"ocker
\etalk1997\nocite{bloecker:96b}; Herwig
\etalk1997,\,1998,\,1999b\nocite{herwig:97,herwig:98b,herwig:98c}).

We give a brief review of AGB star properties relevant for our new
models (\kap{sec:properties}), some remarks on the stellar evolution
code as well as some information on overshoot and its treatment in our
models (\kap{sec:code+overshoot}).  We explain the differences between
models without overshoot and with overshoot by looking separately at
the two relevant convective boundaries: the bottom of the envelope
convection (\kap{sec:overshoot-at-envelope-bottom}) and the bottom of
the PDCZ (\kap{sec:overshoot-at-pdczb}).  The chemical and structural
surface properties of models with overshoot are described in
\kap{sec:3msun-models}. Conclusions are presented in
\kap{sec:conclusions}.


\section{Some properties of AGB stars, stellar models, and overshoot}
\label{sec:properties}

\paragraph{Carbon stars and the third \dup}
AGB stars show a large range of surface C/O ratios from about the
solar value $ \simeq 0.4$ up to well above unity.  In particular many
low-mass AGB stars show up as carbon stars with $\mem{C}/\mem{O} > 1$
\cite{smith:87,frogel:90}. Half of all planetary nebulae are carbon
rich as well \cite{zuckermann:86}.  Already the early stellar
evolution calculations \cite{iben:75,iben:77,iben:78,sackmann:80}
pointed towards the TDUP to provide a link between the intershell
region where carbon is present and the bottom of the envelope
convection.  \footnote{The third \dup\ is to be distinguished from the
  \emph{first} \dup\ which occurs during the first ascent of the giant
  branch after central hydrogen exhaustion and the \emph{second} \dup\ 
  which occurs during the early AGB phase for more massive AGB stars
  after the end of core helium burning.}  However, these models have
not shown sufficient \dup\ for AGB models with low (core) masses.
Lattanzio \cite*{lattanzio:89a} used a method to determine the
convective boundary which ensures that the ratio of radiative and
adiabatic temperature gradient approaches unity smoothly at the
convective boundary. This method favors the occurrence of the TDUP
even for low mass stars of solar metallicities. Hollowell \& Iben
\cite*{hollowell:88} found for low mass and low metallicity that the
\dup\ of processed material occurs if some additional mixing is
assumed. Wood \cite*{wood:81} and Boothroyd \& Sackmann
\cite*{boothroyd:88} showed that lower metallicity and increased
mixing length parameter enhances TDUP.  Straniero
\etal\,\cite*{straniero:97} find the TDUP for low-mass AGB stars and
relate its occurrence to increased numerical resolution.  Calculations
which do not find \dup\ for low core masses are the rule rather than
the exception
\cite{vassiliades:93,bloecker:95a,forestini:97,wagenhuber:98,langer:99}.
Herwig \etal \cite*{herwig:97} presented first results of a 3\msun\ 
stellar model sequence which had been computed with convective
overshoot, treated time-dependently at all convective boundaries.
With this approach TDUP was very efficient for low core masses and
solar metallicity.

\paragraph{Intershell abundances}
The abundance distribution in the intershell is of great importance
because in this region the major nuclear burning and mixing processes
associated with the He-flash and the TDUP take place.  The intershell
abundances are determined by nuclear burning during the quiescent
interpulse phase (mainly hydrogen burning), the nuclear processing
during the TP, the convective mixing in the pulse-driven convective
zone, and third dredge-up. The intershell abundance is significantly
affected if overshoot is applied to the PDCZ because additional
material is mixed into the intershell from the C/O core
(\kap{abundance-intershell-overshoot}).

It affects the local nuclear production, the surface enrichment and
also the structure of the star. It is also of relevance for the
interpretation of surface abundances of hydrogen-deficient post-AGB
stars (see e.g.\ Sch\"onberner,\,1996\nocite{schoenberner:96}; Iben
\etalk1996\nocite{iben:96} or Werner \etalk1999\nocite{werner:98}).

Neither Sch\"onberner \cite*{schoenberner:79} nor Boothroyd \&
Sackmann \cite*{boothroyd:88c} considered overshoot and both found
that the abundance distribution in the intershell evolves with each
TP. Starting with an almost pure He composition from previous hydrogen
burning at the first TP, the abundances approach typical values of
(He/C/O)=(0.76/0.22/0.02) (mass fractions) after a few TPs.

\paragraph{\cdr\ production}
The observed correlation of \spr\ elements and carbon in low-mass
stars \cite{smith:90} points to low-mass stars as a likely site for
n-capture nucleosynthesis \cite{gallino:97a,wallerstein:97}.  Gallino
\etal\cite*{gallino:97b} have demonstrated that, \emph{if} hydrogen
ingestion from the envelope into the intershell region at the end of
the TDUP is assumed \emph{then} the subsequent formation of \cdr\ via
the nuclear reaction $\czw(\p,\gamma)\ndr(e^+\nu)\cdr$ does indeed
predict a neutron exposure which leads to a \spr\ enhancement in
compliance with the solar main component of heavy elements. However,
the physical mechanism of H-ingestion is unclear. Depth-dependent
overshoot naturally predicts the formation of a \cdr\ pocket as needed
for the \spr\ nucleosynthesis.

\paragraph{Mixing and overshoot}
Mixing of elements in stars is attributed to a number of processes
\cite{pinsonneault:97} of which convection is the most effective.
Stellar evolution models commonly employ the mixing-length theory
(MLT) \cite{boehm-vitense:58} or some descendent thereof.  The
boundary of convective instability is determined by the local
Schwarzschild condition
\begin{math}
\label{schwarzschild}
\gradrad > \gradad
\end{math}
where \gradad\ and \gradrad\ are the adiabatic and radiative gradient
\cite{kippenhahn:90}.  However, neighboring layers are related by
inertia, momentum transfer and the equation of continuity and
therefore convective elements might overshoot beyond the boundary of
convection (e.g. Shaviv \& Salpeter,\,1973\nocite{shaviv:73};
Maeder,\,1975\nocite{maeder:75}; Roxburgh,\,1978\nocite{roxburgh:78};
Bressan \etalk1981\nocite{bressan:81};
Langer,\,1986\nocite{langer:86}).  No commonly accepted quantitative
theoretical description of convective overshoot currently exists
\cite{renzini:87}.

Comparison of stellar models with observational findings indicates
that convective overshoot occurs in real stars
\cite{mermilliod:86,maeder:89,andersen:90,stothers:91,napiwotzki:91,alongi:91,alongi:93,schroeder:97,kozhurina:97}.
Overshooting in stellar evolution calculations has been simulated by
an extension of the instantaneous mixing beyond the convective
boundary (instantaneous overshoot). The widely used set of models by
Schaller \etal\cite*{schaller:92} has been calculated with an
overshoot distance of one fifth pressure scale height
($\alpha_{over}=0.2$) which was calibrated in order to fit the
observed terminal age \ms\ of 65 stellar clusters and associations.
Improving on the instantaneous treatment, Deng
\etal\cite*{deng:96a,deng:96b} and Salasnich
\etal\cite*{salasnich:99}) have explored the effects of turbulent
diffusion in the overshoot region of massive stars.

Due to the insights of two- as well as three-dimensional
hydrodynamical simulations convection is nowadays pictured in terms of
downdrafts and up-flows rather than as a hierarchy of eddies
\cite{stein:98}.  Two-dimensional simulations showed that prominent
downward-directed plumes can overshoot a substantial distance into the
stable region \cite{hurlburt:86,hurlburt:94,freytag:96}.  These models
show emerging and vanishing patterns of curls, fast narrow downdrafts
and broad up-flow regions.The turbulent velocity field decays
exponentially beyond the convective boundary (see also
Xiong,\,1985\nocite{xiong:85}; Asida \& Arnett,\,in prep.).  These
results have been applied via a time-dependent treatment of convective
mixing to low- and intermediate mass stellar models by Herwig \etal
\cite*{herwig:97,herwig:00c} (low mass TP-AGB), \mbox{Ventura \etal
  \cite*{ventura:98}} (main sequence) and Mazzitelli
\etal\cite*{mazzitelli:99} and Bl\"ocker \etal\cite*{bloecker:99c}
(massive AGB stars, HBB).

\section{The stellar evolution code and computations}
\label{sec:code+overshoot}

\paragraph{General aspects of the code}
The results presented in this paper have been obtained with the
stellar evolution code described by Bl\"ocker \cite*{bloecker:95a}
with the modifications given in Herwig \etal \cite*{herwig:97}.  Here,
we will add only a few relevant details.  The
$\czw(\alpha,\gamma)\ose$ reaction rate has been taken from Caughlan
\& Fowler \cite*{caughlan:88} and multiplied by a factor of $1.7$ as
recommended by Weaver \& Woosley \cite*{weaver:93}.  The reaction
rates $\osi(\p,\gamma)\fac$ and $\osi(p,\alpha)\nvi$ have been taken
from the compilation of Landr\'e \etal \cite*{landre:90} as described
in El Eid \cite*{eleid:94}. The model sequence without overshoot has
been computed with the latest \emph{OPAL} opacity tables
\cite{iglesias:96}. This choice had purely technical reasons but does
not affect the comparison.

The adjustment of the numerical resolution (geometry and time) plays
an important role in the computation of AGB models
\cite{straniero:97,frost:96,mowlavi:99}. The resolution is adjusted
from model to model according to the changes of the state variables as
a function of time and location within the star. For example, the
luminosity due to nuclear processing of helium \lhe\ should not change
more than $5\%$ between two models. The models have $\simeq 1900 \dots
2300$ mass grid points which accumulate around the core-envelope
interface where the pressure and density drop steeply. The model time
steps range from about a dozen years during the quiescent H-burning
interpulse phase down to typically a few days during the He-flash and
the following phase where \dup\ may occur (\kap{sec:tdup}).

\paragraph{Exponential diffusive overshoot}
For the time-dependent treatment of overshoot mixing we follow the
prescription of Freytag \etal \cite*{freytag:96}.  The particle
spreading in the overshoot region can be described as a diffusion
process and the diffusion coefficient $D_\mem{OV}$ can be fitted by a
formula like \beq D_\mem{OV}(z) = t_\mem{c}\cdot v^2_\mem{rms}(z) \eeq
where $z$ denotes the geometric distance from the edge of the
convective zone and $t_\mem{c}$ is a characteristic time scale for the
considered convection zone.  Combining this with the finding of an
exponential decay of the velocity field, Freytag \etal
\cite*{freytag:96} give the diffusion coefficient in the overshoot
region as

\begin{equation} \label{dhyd}
           D_{\rm OV} = D_0 \exp{\left( \frac{-2 z}{H_{\rm v}}\right)}, \,\,\,\,\,\,\,
           H_{\rm v} = f \cdot H_{\rm p},
\end{equation}
where $H_{\rm v}$ is the velocity scale height of the overshoot
convective elements at the convective boundary. $H_{\rm v}$ can be
expressed as a fraction $f$ of the pressure scale height $H_{\rm p}$.
The quantity $D_{0}$ is the expression $v^2_\mem{edge}t_\mem{c}$ in
Freytag et al.\ (1996: Eq.\,9). Here we have approximated $D_{0}$ by
the diffusion coefficient in the convectively unstable region near the
convective boundary $r_{\rm edge}$. Note that $D_{0}$ is well defined
because $D_\mem{c}$ approaches the convective boundary with a small
slope and drops almost discontinuously at $r_{\rm edge}$.  The
diffusion coefficient $D_\mem{c}$ in the convection zone can be
derived from the MLT (see Langer \etalk1985\nocite{langer:85}).  In
the radiative zone where the diffusion coefficient has dropped below
$D_{\rm OV}^{\rm limit}=\nat{1}{-2}\mem{cm^2/s}$ no element mixing is
allowed ($D_{\rm rad}=0$).

\begin{figure*}[hbtp] 
  \epsfxsize=\textwidth
  \epsfbox{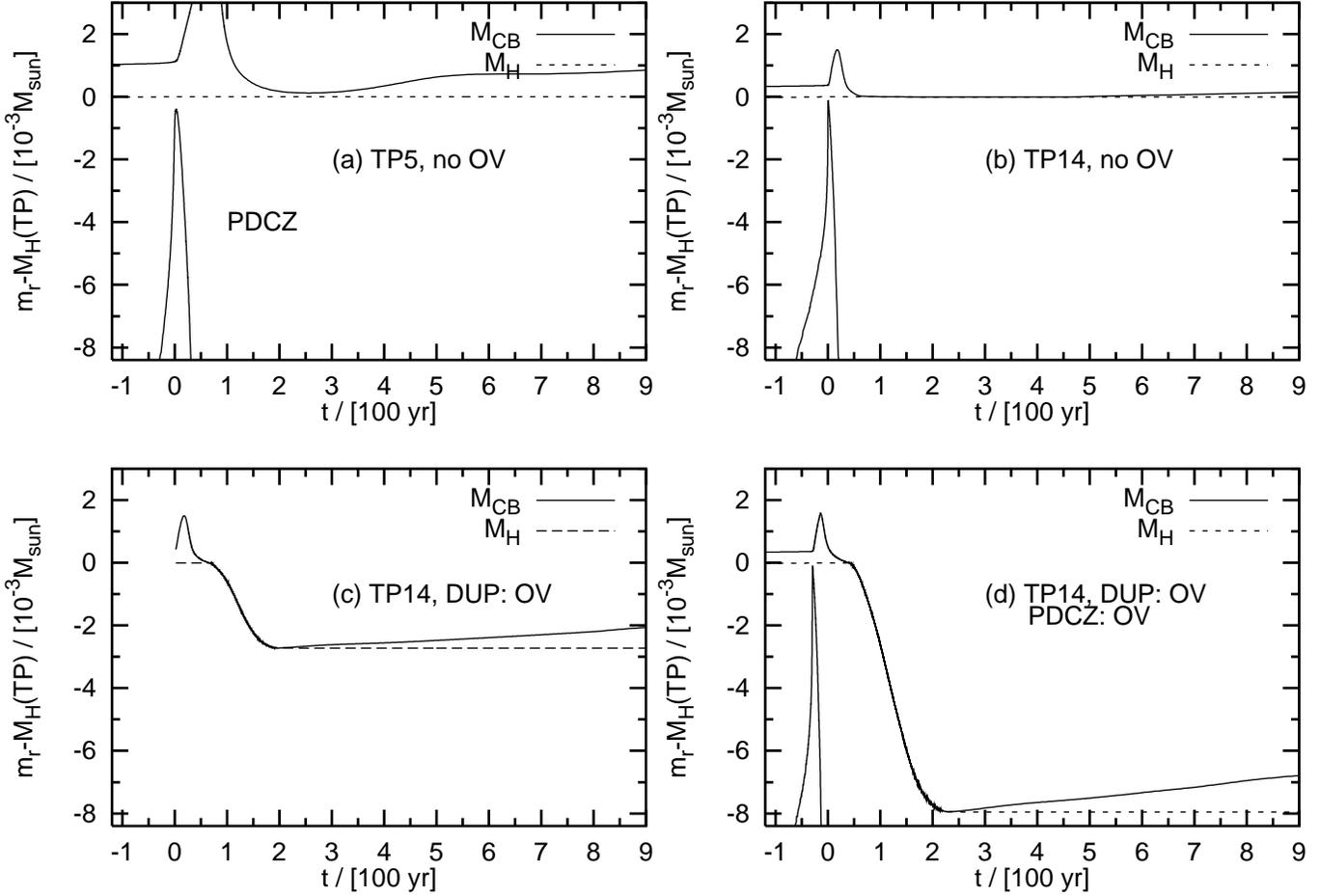}
\caption[]{ \label{dup-all} 
  Evolution of mass coordinate of convective boundary
  ($\mathsf{M_{CB}}$, solid line) and H-free core ($\mathsf{M_{H}}$,
  dashed line). The mass scale has been set to zero at the mass
  coordinate of the H-free core at the time of the He-flash (TP5:
  $\mh=0.6421\msun$; TP14: $\mh=0.6917\msun$). The time scale has bee
  set to zero at the peak of the He-flash luminosity (TP5:
  $t_\mem{0}=227034\jahre$; TP14: $t_\mem{0}=677076\jahre$).  The
  spike (solid line) at $t=0\jahre$ shown in (a), (b) and (d) is the
  top of the pulse-driven convection zone of the He-flash. The upper
  solid line shows the bottom boundary of the envelope convection
  zone.  Panel (a): 5$^\mem{th}$ TP of the 3\msun\ sequence without
  overshoot; panel (b): 14$^\mem{th}$ TP of the same sequence; panel
  (c): the same TP as in (b) started after the PDCZ has vanished and
  with overshoot at the bottom of the envelope convection zone; panel
  (d): again the 14$^\mem{th}$ TP recalculated from before the onset
  of the He-flash, with overshoot at the PDCZ and during the following
  TDUP episode.}
\end{figure*}
The free parameter $f$ in \gl{dhyd} describes the efficiency of the
extra diffusive mixing. For larger $f$ the extra partial mixing beyond
the convective edge extends further.  In their simulations Freytag
\etal\cite*{freytag:96} found velocity scale heights of the order of a
pressure scale height ($f \simeq 1$).  However, under adiabatic
conditions the convective flows move so fast that they will not be
able to exchange heat efficiently with the surrounding when entering
into the stable region. Accordingly a stronger deceleration of the
plumes is likely in the stellar interior compared to the situation of
the shallow surface convection zones. In order to obtain an order of
magnitude for $f$ in the stellar interior it has been scaled to
reproduce the observed width of the main sequence. A parameter of $f
\simeq 0.016$ was found to reproduce the models of Schaller
\etal\cite*{schaller:92} (see end of \kap{sec:properties}).  Using AGB
overshoot models as a starting point for post-AGB models of
H-deficient objects has confirmed this assumption a posteriori, at
least for the bottom of the He-flash convection zone
\cite{herwig:99c}.

The overshoot prescription (\gl{dhyd}) has been applied irrespective
of $\mu$-gradients which may decrease the overshoot efficiency
\cite{langer:83}.  The analysis of old open clusters even suggests
that overshoot may not expand against $\mu$-gradients in low mass
stars \cite{aparicio:91}. Theoretically, Canuto \cite*{canuto:98a} has
recently argued that $\mu$-gradients do decrease the overshoot
efficiency, but do not prevent the effect entirely. In our study, this
aspect becomes important at the bottom of the envelope convection
during the third dredge-up. As explained in \kap{sec:oversh-dredge}
the results are not significantly dependent on the amount of overshoot
at this convective boundary.

\begin{figure}[thbp] 
  \epsfxsize=8.8cm \epsfbox{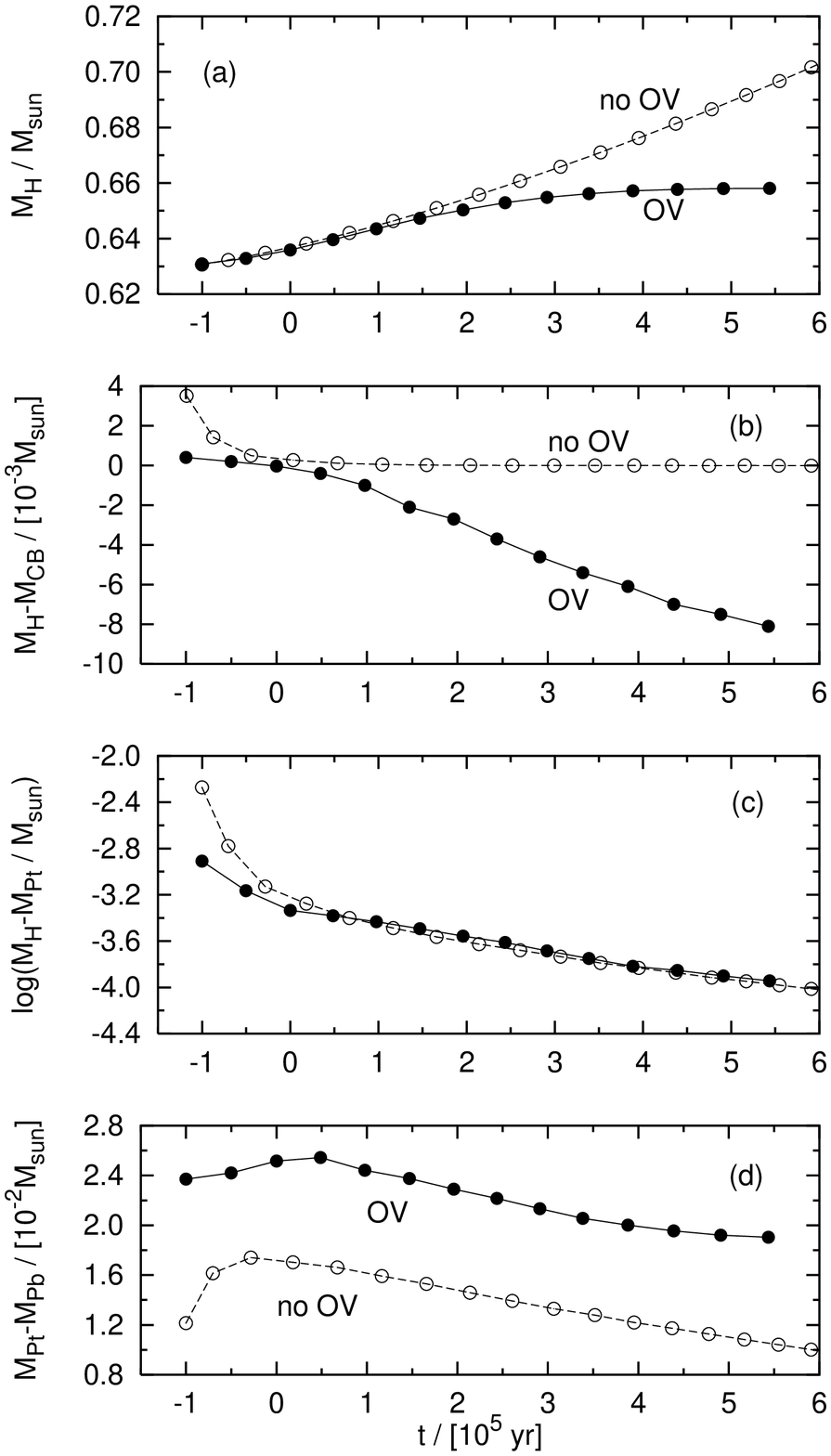}
\caption[]
{ \label{fig:mass-all} Evolution of the position and size in mass
  coordinates of several quantities at each TP as a function of time.
  Full symbols show the 3\msun\ sequence computed with overshoot at
  all convective boundaries, open symbols represent the comparison
  sequence without any overshoot and started with the same initial
  model before the first TP as the 3\msun\ sequence with overshoot.
  Panel (a): H-free core at the TP, the increase by H-burning during
  the interpulse period is counterbalanced by the dredge-up after the
  TP for the overshoot sequence; (b): difference of core mass at TP
  and smallest mass coordinate of convective envelope bottom achieved
  after the TP, negative values indicate DUP; (c): difference of core
  mass at TP and the largest mass coordinate of the PDCZ top achieved
  during the TP, (d): difference of top and bottom mass coordinate of
  PDCZ.  }
\end{figure}

\begin{figure}[t] 
  \epsfxsize=8.8cm \epsfbox{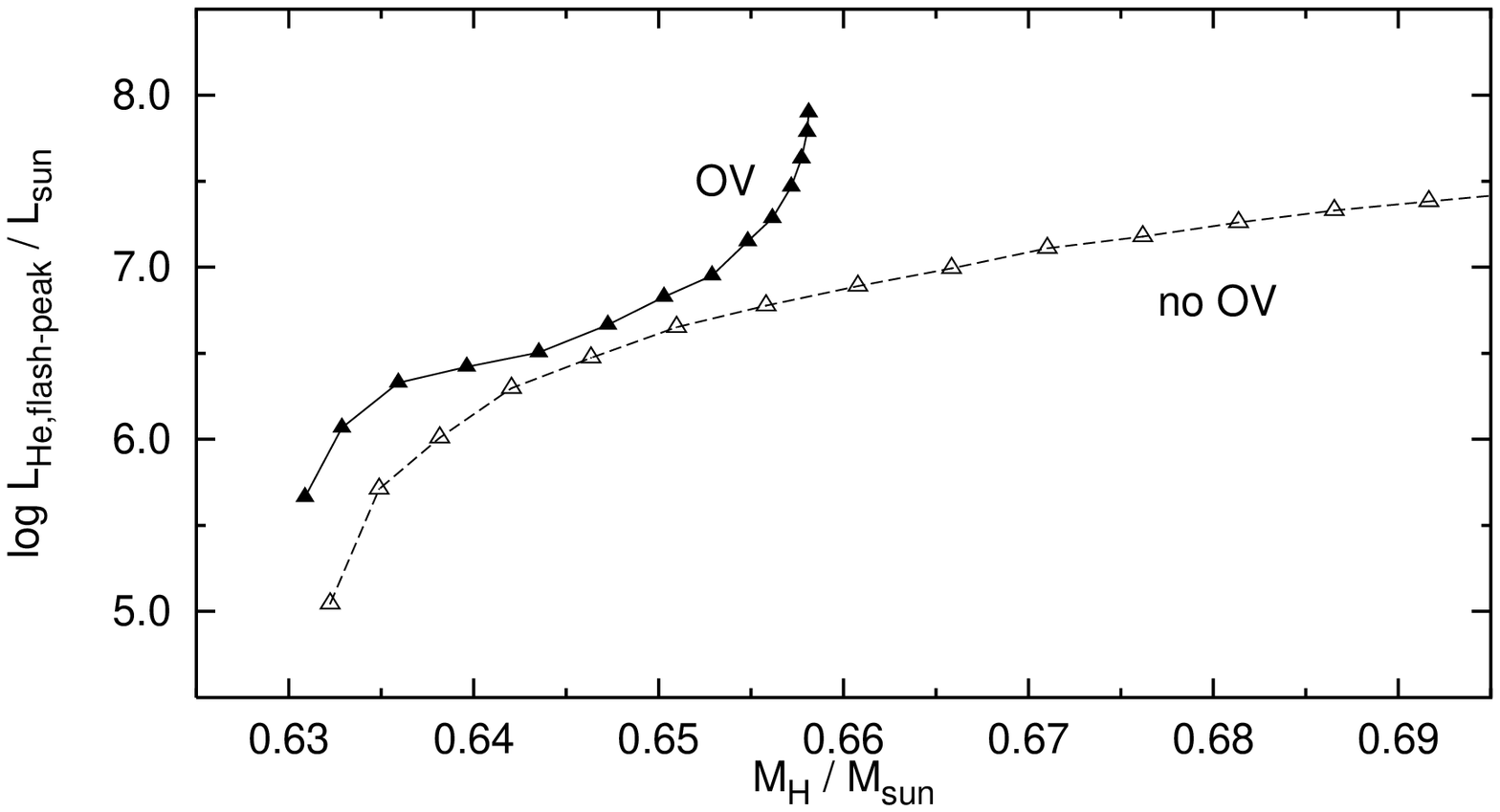}
\caption[]
{ \label{fig:Mc-LHe} The maximum helium-burning luminosities during
  the He-flash for the 3\msun\ model sequence with (filled triangles)
  and without (open triangles) overshoot. Each symbol corresponds to
  one TP.  The generally larger peak flash luminosities of the
  overshoot TPs are due to the extra-mixing below the bottom of the
  He-flash convection zone which leads to higher temperatures in the
  He-shell during the flash (\abb{fig:TP5-STRUC-all}). The relation
  runs vertically at $\mh \simeq 0.657\msun$ for the OV case because
  efficient dredge-up prevents further core mass growth ($\lambda
  \simeq 1$).  }
\end{figure}

\begin{figure}[tph] 
  \epsfxsize=8.8cm \epsfbox{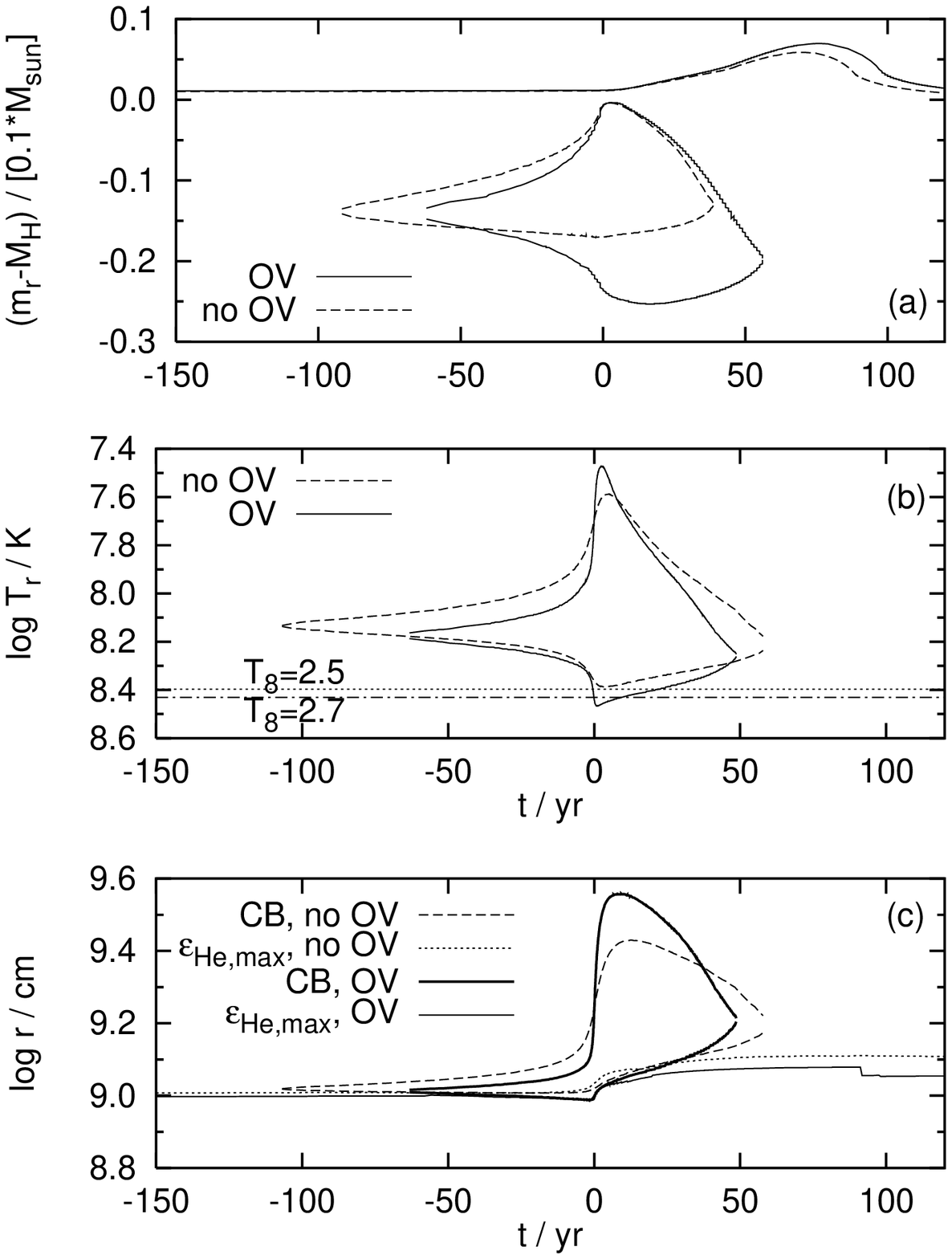}
\caption[]
{ \label{fig:TP5-STRUC-all} Panel (a): The mass coordinates of the
  PDCZ boundary during the fifth TP of the 3.0\msun\ with and without
  overshoot.  The time has been set to zero at the moment where the
  He-burning luminosity has reached the maximum.  Panel (b):
  Temperature at the boundary of the He-flash convection zone of the
  same TP.  Panel (c): Geometric position of the PDCZ boundary (CB)
  and of the maximum energy release by helium burning.  }
\end{figure}

\begin{figure}
  \epsfxsize=8.8cm \epsfbox{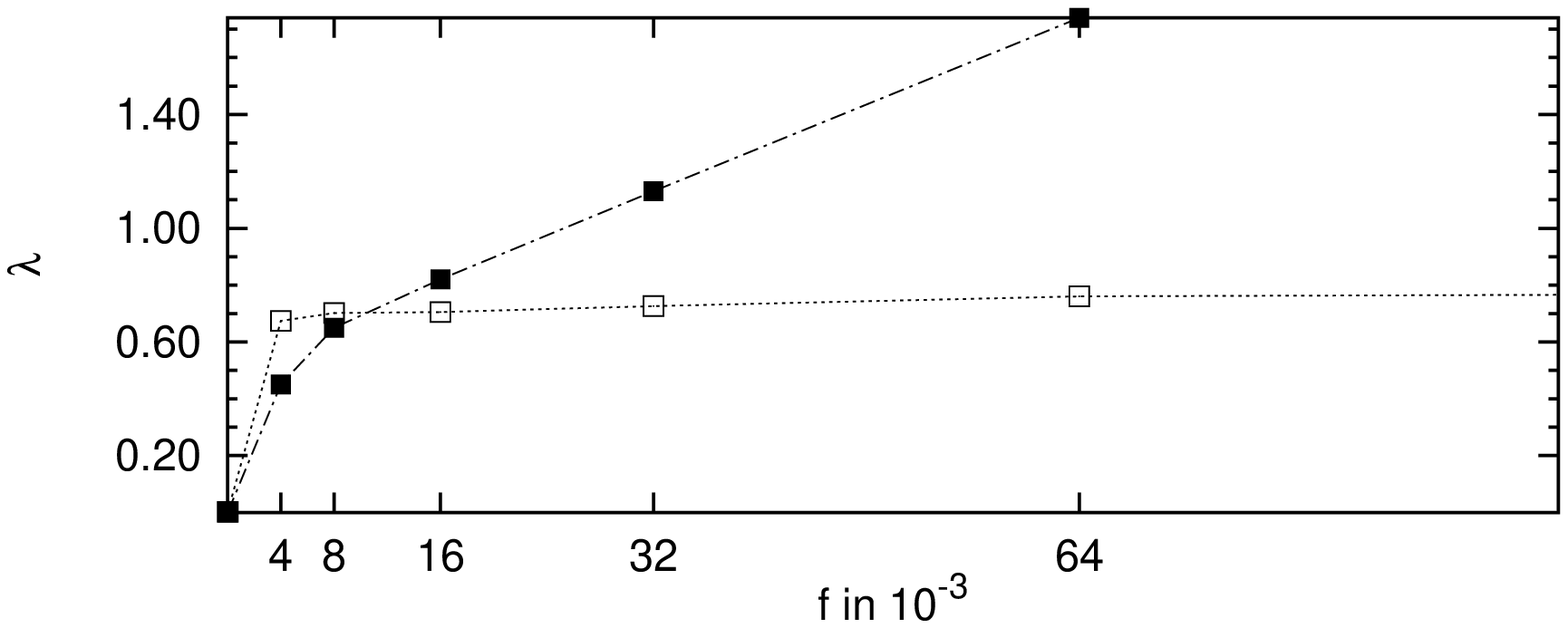}
\caption{ \label{fig:f-lambda} 
  Variation of the dredge-up efficiency $\lambda$ with the overshoot
  efficiency. The open symbols represent cases where the $f$ value has
  been changed just after the \pdcz\ has disappeared after the eighth
  TP but before the bottom of the envelope convection has come close
  to the mass coordinate of the H-free core. Thus, in this case a
  different overshoot parameter is only effective at the bottom of the
  envelope during the dredge-up phase. The full symbols show the
  dredge-up efficiency after the following ninth TP which has been
  computed with the respective $f$-value also for the He-flash
  convection zone.  In this case a different overshoot parameter has
  been effective at the bottom of the He-flash convection zone.}
\end{figure}

\begin{figure}[t] 
  \epsfxsize=8.8cm \epsfbox{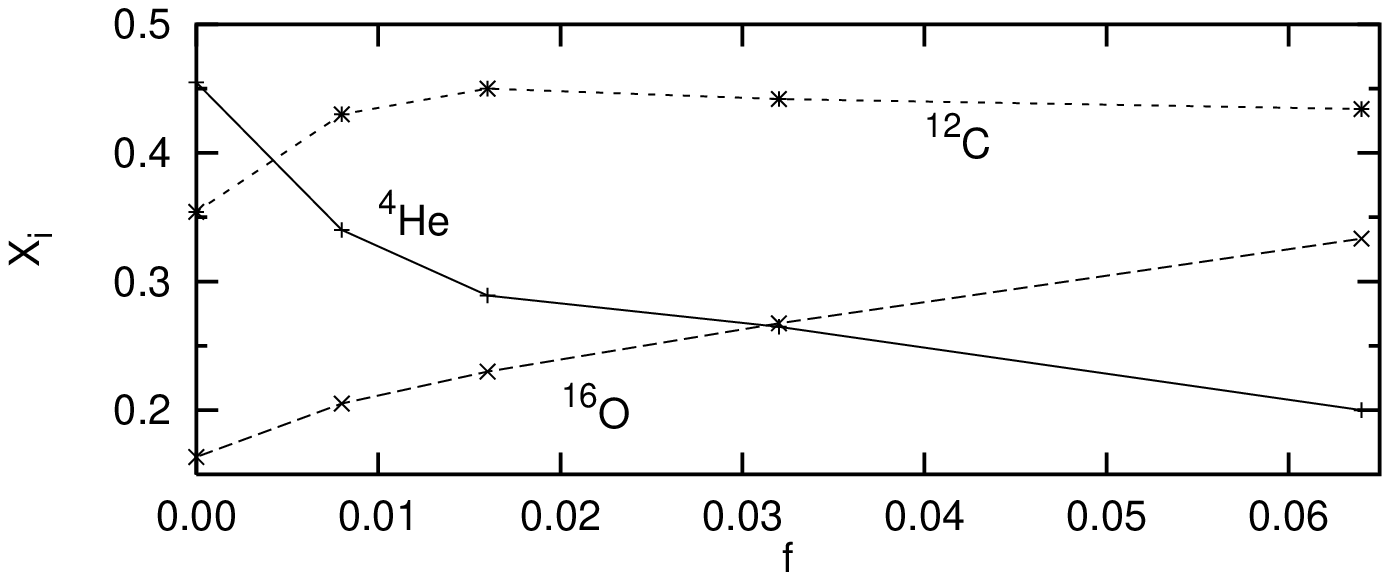}
\caption[]
{ \label{f-ABUND-TP8} The intershell abundances after the ninth TP as
  a function of $f$. While for \abb{f-Xi-TP3} the different overshoot
  values have been applied over all previous TPs here we have computed
  only the ninth TP with different overshoot values. The starting
  model has been a 3\msun, $f=0.016$ model before the ninth TP. This
  figure demonstrates how the abundances are modified due to overshoot
  during only one TP.  }
\end{figure}

\begin{figure}[t] 
  \epsfxsize=8.8cm \epsfbox{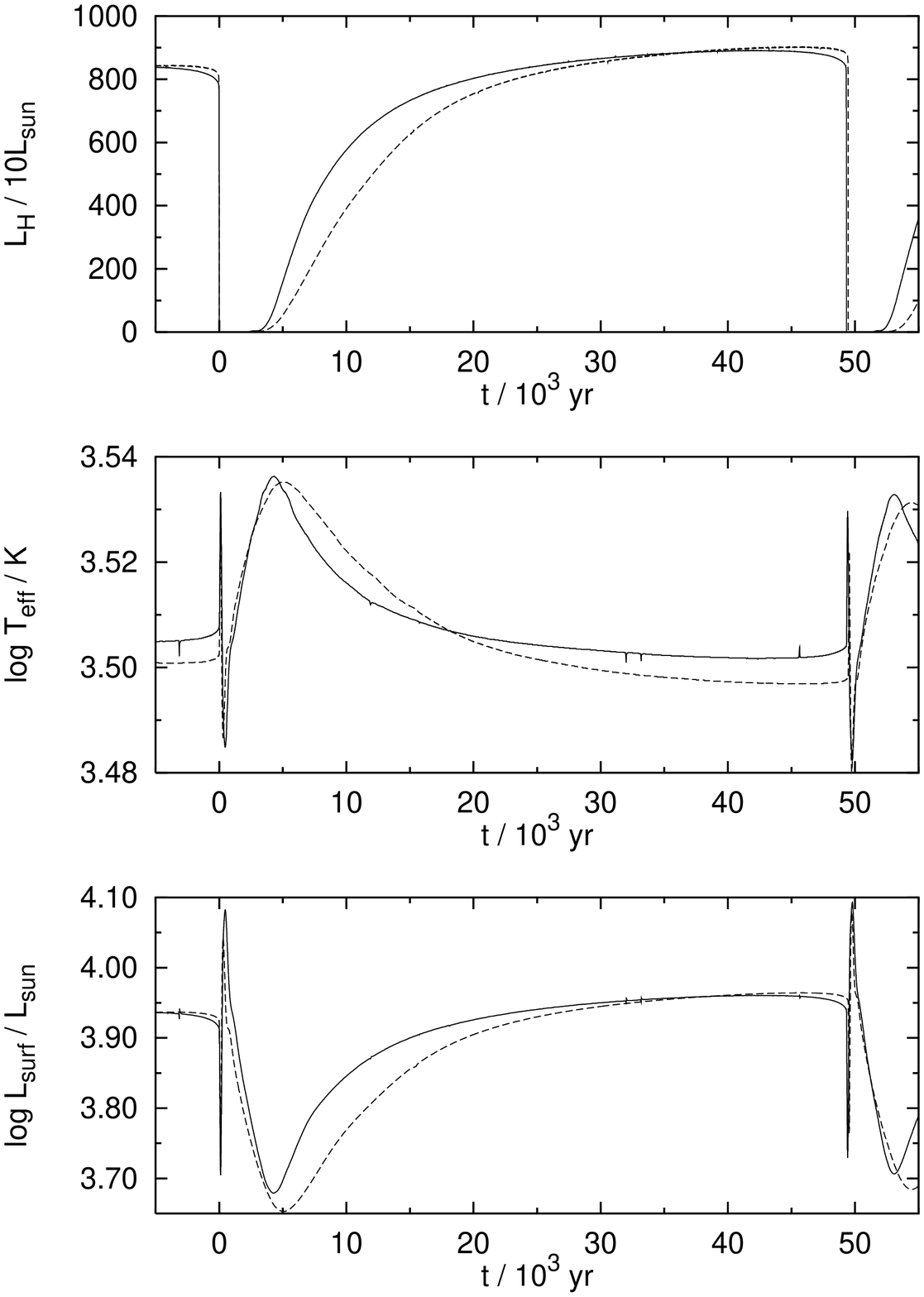}
\caption[]
{ \label{fig:LSURF} Luminosity of H-burning (top panel), effective
  temperature (middle panel) and surface luminosity (bottom panel) for
  one complete pulse cycle of the sequence with (solid line) and
  without (dashed line) overshoot respectively. Time has been set to
  zero at the fifth TP.  }
\end{figure}

\paragraph{The stellar evolution computations}
The findings in this work are based on two model sequences of 3\msun\ 
and 4\msun\ with overshoot. For comparison, one 3\msun\ TP-AGB
sequence has been computed without overshoot. It has been started from
a 3\msun\ overshoot model taken just before the occurrence of the
first TP (Fig.\,\ref{fig:mass-all}, \ref{fig:Mc-LHe},
\ref{fig:TP5-STRUC-all}, \ref{fig:LSURF}). In order to study the
dependence of the evolution of the intershell abundance as a function
of the parameter $f$ a few more sequences with different $f$-values
have been computed from this starting model (\abb{f-Xi-TP3}) and
followed over three thermal pulses. In addition, the eighth TP cycle
of the 3\msun\ overshoot case has been recomputed for different $f$
values (Fig.\,\ref{fig:f-lambda}, \ref{f-ABUND-TP8}). In \abb{dup-all}
a recomputation of the 14$^\mem{th}$ TP of the 3\msun\ sequence
initially without overshoot is shown.

\begin{figure}[tbhp] 
  \resizebox{\hsize}{!}{\includegraphics{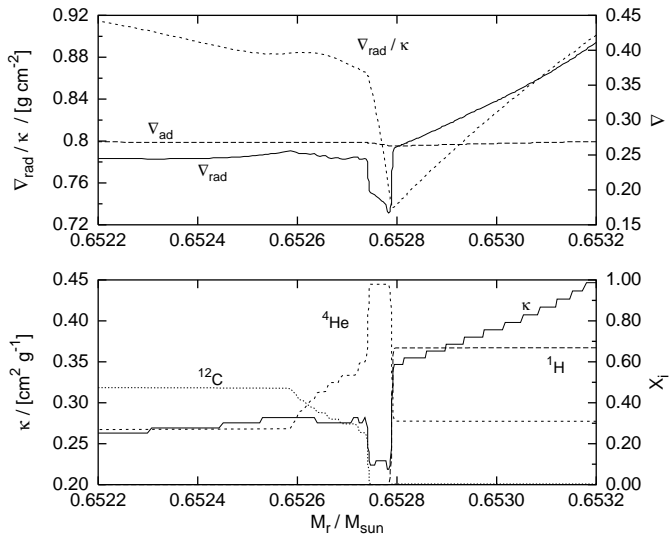}}
  \caption[Temperature gradients, opacity and abundances 
  during the TDUP ]{ \label{RAD-TP8.53899} Temperature gradients,
    opacity and abundances for the mass fraction in the AGB star
    during the TDUP ($f=0.016$) after the eighth TP with overshoot.
    $t=216\jahre$ after peak helium luminosity. Compare
    \abb{DEPSf0.0}: there the bottom of the convective envelope does
    not reach into the He-rich buffer zone at $M_\mem{r}=0.6529\msun$,
    whereas in the case shown here the envelope has started to engulf
    the He-rich buffer zone.  }
\end{figure} 
A solar-like initial composition, $(Y,Z)=(0.28,0.02)$, has been chosen
corresponding to Anders \& Grevesse \cite*{anders:89}. The
mixing-length parameter of the MLT is $\alpha_\mem{MLT} = 1.7$ which
has been calibrated by reproduction of the solar surface parameters. A
Reimers-type mass loss with efficiency $\eta=1$ has been applied. The
Reimers mass loss formula has not been designed for AGB stars, however
for the purpose of this study mass loss is not important. The 3\msun\ 
(4\msun) model sequence has a total mass of 2.686\msun\ (3.689\msun)
at the 12$^\mem{th}$ (11$^\mem{th}$) TP (see \tab{3Meta1f} and
\ref{4Meta1f}).

Overshoot has been applied during all evolutionary stages and to all
convective boundaries. Overshoot applied to the main sequence core
convection has indirect implications for AGB star models as it shifts
the core masses at the first thermal pulse upward and decreases the
progenitor mass ($M_\mem{up}$) limit for C/O white dwarfs
\cite{alongi:93}. If overshoot is also applied during the second DUP
phase this effect is counterbalanced somewhat and $M_\mem{up} \simeq
6.5$ (T. Driebe: priv.\ com., see also Bl\"ocker
\etal\cite*{bloecker:99c} and Weidemann \cite*{weidemann:00}). This is
just above the value obtained from models in which overshoot is only
applied during the \ms\ phase \cite{bressan:93}.

\section{Overshoot at the bottom of the
  convective envelope}
\label{sec:overshoot-at-envelope-bottom}

Overshoot at the bottom of the convective envelope has an effects on
the modeling of the TDUP and on the formation of a \cdr\ pocket after
the end of the \dup\ phase.

\subsection{The third dredge-up}
\label{sec:tdup}

\begin{figure*}[hbtp] 
  \begin{center}
    \epsfxsize=0.8\textwidth \epsfbox{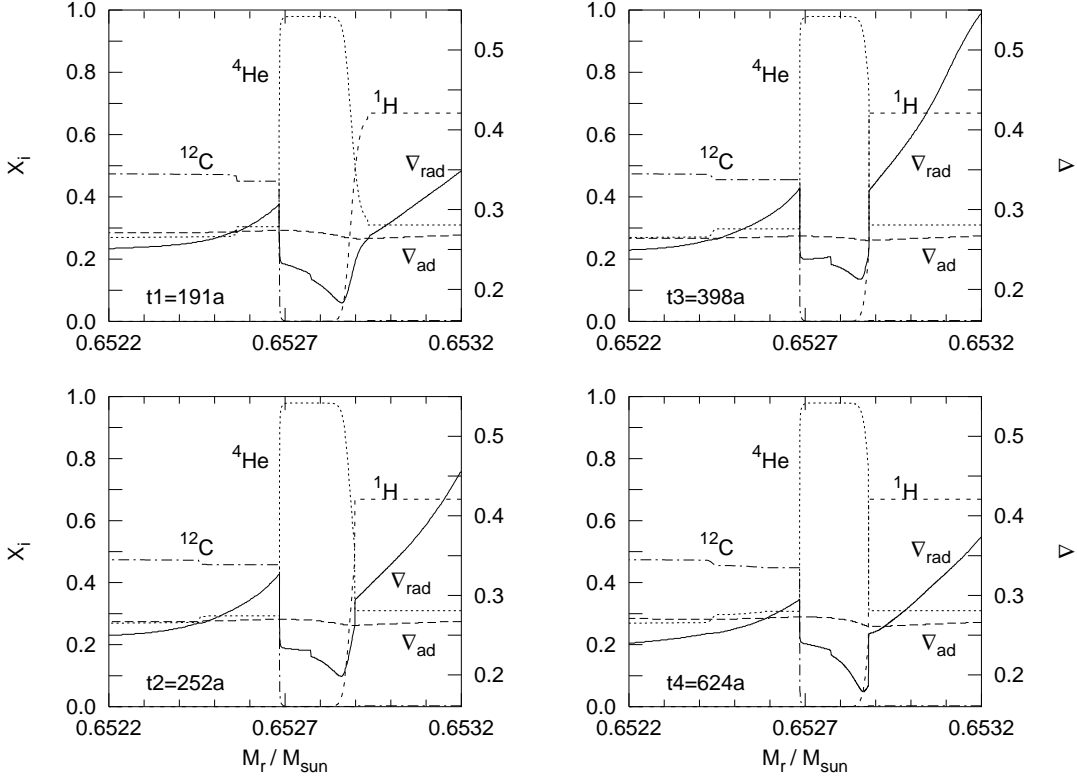}
  \end{center}
  \caption[Abundances
  and radiative and adiabatic gradient ($f=0.0$)]{ \label{DEPSf0.0} A
    time series of the region of the former hydrogen-burning shell
    after the eighth TP of the 3\msun\ sequence. The dredge-up phase
    after the thermal pulse of the overshoot sequence has been
    recalculated without overshoot and as a result dredge-up is
    prevented (compare \abb{dup-all}) and \abb{RAD-TP8.53899}.
    Abundances in mass fractions of \hevi\ (short-dashed), \czw\ 
    (dot-dashed) and hydrogen (dashed) refer to the left scale while
    the adiabatic (\gradad, long-dashed) and radiative (\gradrad,
    solid line) temperature gradient refer to the right scale. For
    each panel the abscissa ranges from the bottom of the envelope at
    the right to the top of the intershell region at the left.  In the
    middle of each panel a region of almost pure \hevi\ has been
    formed by hydrogen-burning during the previous interpulse phase.
    The lower boundary of the He-rich region is formed by the He-flash
    convection zone, which has at $t=0\jahre$ homogenized the
    intershell layers below $M_\mem{r}=0.6527\msun$.  The four
    snapshots show how the bottom of the envelope convection zone
    advances downward by mass during the phase when TDUP would occur
    with overshoot.  The order is from top to bottom and left to right
    and the ages are given with respect to the flash peak of the
    previous pulse.  Note how the discontinuity of \gradrad\ is
    related to the abundance discontinuity which starts to build up
    ($t=252\jahre$). The discontinuity grows further ($t=398\jahre$)
    and finally when the region resumes contraction the bottom of the
    envelope convection retreats without being able to dredge-up
    material from the H-exhausted regions.  }
\end{figure*}

It has been a matter of discussion whether convective overshoot at the
bottom of the convective envelope is a prerequisite for \dup\ 
\cite{mowlavi:99}. This question is closely linked to the general
problem of determining the convective boundary of mixing within the
framework of a local theory of convection together with its numerical
implementation.  As reported by previous authors (e.g.\ 
Paczy\'nski,\,1977\nocite{paczynski:77}), our models without overshoot
develop a discontinuity in \gradrad\ when the bottom of the envelope
convection zone reaches the He-rich buffer layer. This discontinuity
can prohibit the TDUP under certain conditions. In our models of AGB
stars with not so massive cores the flux on the stable side of the
convective boundary is not large enough to lift \gradrad\ above
\gradad.

In \abb{dup-all} the situation in the dredge-up region is shown for
several thermal pulses computed with different assumptions on
overshoot.  After an early TP of the 3\msun\ sequence computed without
any overshoot (panel a) the bottom of the envelope convection does not
approach the H-free core close enough for any dredge-up. For more
advanced TP (panel b) the convection clearly approaches the H-free
core for a period of about $300 \jahre$ but due to the effect of the
abundance discontinuity shown in \abb{DEPSf0.0} the lack of overshoot
prohibits the TDUP. This is evident from panel (c) which shows the
same situation with overshoot switched on below the bottom of the
envelope convection. In this case a considerable dredge-up of material
is possible. Panel (d) shows that overshoot applied also to the
preceding PDCZ increases the DUP efficiency even further, which is
discussed in detail in \kap{sec:overshoot-at-pdczb}.  Note that if
overshoot only at the base of the convective envelope had been applied
for the 5$^\mem{th}$ TP shown in \abb{dup-all}, this would have had no
effect on DUP because the stable layer between the bottom boundary of
convection and the H-free core is too large compared to the tiny
extent of the overshoot layer. \abb{DEPSf0.0} demonstrates the
situation of panel (b) of \abb{dup-all} as a set of profiles. The
bottom of the convective envelope recoils at the He-rich region which
encloses the core, despite the fact that contact between the
convective envelope and the H-free core is established over a few
hundred years. In \abb{DEPSf0.0} the time after the eighth TP of the
3\msun\ sequence with overshoot has been recalculated without
overshoot at the bottom of the envelope convection. A profile at the
beginning of the TDUP of the corresponding original sequence with
overshoot is shown in \abb{RAD-TP8.53899}. Close inspection and
comparison with \abb{DEPSf0.0} shows that $10^{-4}\msun$ of He-rich
material have already been dredged-up.  The additional mixing removes
the sharp abundance discontinuity. This in turn dissolves the sharp
discontinuity in the radiative gradient \gradrad\ which sensitively
depends on the opacity.  In the envelope, hydrogen has a mass fraction
of $\simeq 0.70$ and a higher opacity than helium. If this material is
mixed into the stable region the radiative gradient will be lifted and
dredge-up can occur.

This principle is not dependent on the efficiency of overshoot within
a factor of maybe two (\abb{fig:f-lambda}).  At the lower edge of the
AGB envelope the convective velocities are of the order of $\simeq 1
\mathrm{km/s}$ while very efficient mixing at this convective boundary
is still achieved by convective velocities down to $\leq 1
\mathrm{cm/s}$.  An instantaneously mixed region can be identified in
which the downward directed flows decelerate from the initial velocity
at the edge to the much lower velocity where the chemical abundance is
not homogenized. Here, the turbulent velocity is so slow that mixing
becomes inefficient.  If the overshoot efficiency is small the region
of homogenized composition may be very thin. This will still suffice
to allow TDUP if the resolution is chosen such that overshoot is
numerically able to smear out the abundance discontinuity (see below).
\abb{RAD-TP8.53899} gives an impression of the masswise size of the
overshoot region which can be barely identified in the abundance
profiles on this scale.  If the time steps are too large with a small
overshoot, the smoothing effect on the abundance discontinuity is not
efficient and DUP is prevented as is the case in the examples shown in
Fig.\,\ref{dup-all}b and \abb{DEPSf0.0}.

In our calculations without overshoot \gradrad\ and \gradad\ are first
compared at each mass grid zone which defines stable and unstable
grids. If a discontinuity has already built up, grids are included at
the discontinuity in order to still resolve the abundance gradient. No
mixing is allowed from an unstable grid to a stable grid. This
approach closely realizes the MLT which predicts that no convective
bubble can leave the convectively unstable region.  However, one can
argue equally well that the average of the gradients between
neighboring grids should be compared. This criterion would allow
material in stable grids to be mixed with material from the
neighboring grid if the latter is unstable. Under certain conditions
dredge-up may occur with the second treatment which would not occur
with the first treatment.

During the \dup\ the upper part of the intershell region is engulfed
by the convective envelope at the rate $\mdot_\mathrm{DUP} = \partial
m_\mem{edge} / \partial t$ where $m_\mem{edge}$ is the mass coordinate
of the convective boundary.  Then, the extent $\Delta M_\mathrm{h}$
(in Lagrangian coordinates) of the stable layer which is practically
homogenized by extra mixing, depends on the overshoot efficiency and
the rate $\mdot_\mathrm{DUP}$.  $\Delta M_\mathrm{h}$ is larger for a
larger efficiency. Moreover, $\Delta M_\mathrm{h}$ will be larger if
$\mdot_\mathrm{DUP}$ is smaller and vice versa because element mixing
is treated time dependently.  With the efficiency $f=0.016$ and
$\mdot_\mathrm{DUP} \simeq \nat{2}{-5} \msun/\jahre$ we find $\Delta
M_\mathrm{h} \simeq 10^{-6}\msun$. In order to resolve the overshoot
region the mesh size at and around the convective boundary should not
exceed $\Delta M_\mathrm{h} / 10 $.

The time step is limited by the extent of the homogenized stable
layer. Current stellar evolution calculations typically separate the
solution of the structural equations and the solution of the mixing
equations (often a diffusion-like equation) in order to save
computational resources. This separated solution is justified if the
two processes described by the two sets of equations operate on
different time scales.  If the numerical time step is of the order
$\Delta t > \Delta M_\mathrm{h} / \mdot_\mathrm{DUP}$ then the mass
coordinate of the envelope convection bottom $m_\mem{edge}$ after this
time step would be below the previously homogenized layer:
$m_\mem{edge}(t+\Delta t) < m_\mem{edge}(t) - \Delta M_\mathrm{h}$.
In that case the separated solution of the structure equation and the
equations of abundance change is no longer valid because the
abundances $ Y(t + \Delta t)$ at $m_\mem{edge}(t+\Delta t)$ cannot be
approximated by the abundances $ Y(t)$. Only if this approximation is
valid can the separate computation of structure and abundance change
give correct results. This is only the case if $\Delta t < \Delta
M_\mathrm{h} / \mdot_\mathrm{DUP}$.

The actual spatial resolution is controlled by a combination of
criteria. Throughout the star we insert mass shells if any of the
conditions $\Delta \log L > 0.01$, $\Delta \log P > 0.02$, $\Delta
\log X(\czw) > 0.02$ or $\Delta \log X(\hevi) > 0.02$ is fulfilled
between two mass shells. After the He-flash, when the \dup\ is
expected we increase the resolution in a region of about $0.06\cdot
\mathrm{M}_\mathrm{core}$ around the mass coordinate at the bottom of
the envelope convection zone by requiring $\Delta \log P < 0.01$. This
prescription for the spatial resolution ensures the above mentioned
mesh size constraints in an adaptive way.  Any further increase of
resolution does not affect the amount of \dup\ obtained. Adapting the
mesh may cause additional numerical diffusion. In the situation of the
advancing convective boundary associated with the TDUP this does not
prevent the discontinuity (\abb{DEPSf0.0}) if no minimum grid size is
enforced. However, numerical diffusion may easily distort the
abundance profiles in the \cdr\ pocket (\abb{C13I}) when the grid
density after the end of the TDUP is first reduced and then increases
afterwards when H-burning starts.

\begin{figure*}[htbp] 
  \epsfxsize=13.5cm \epsfbox{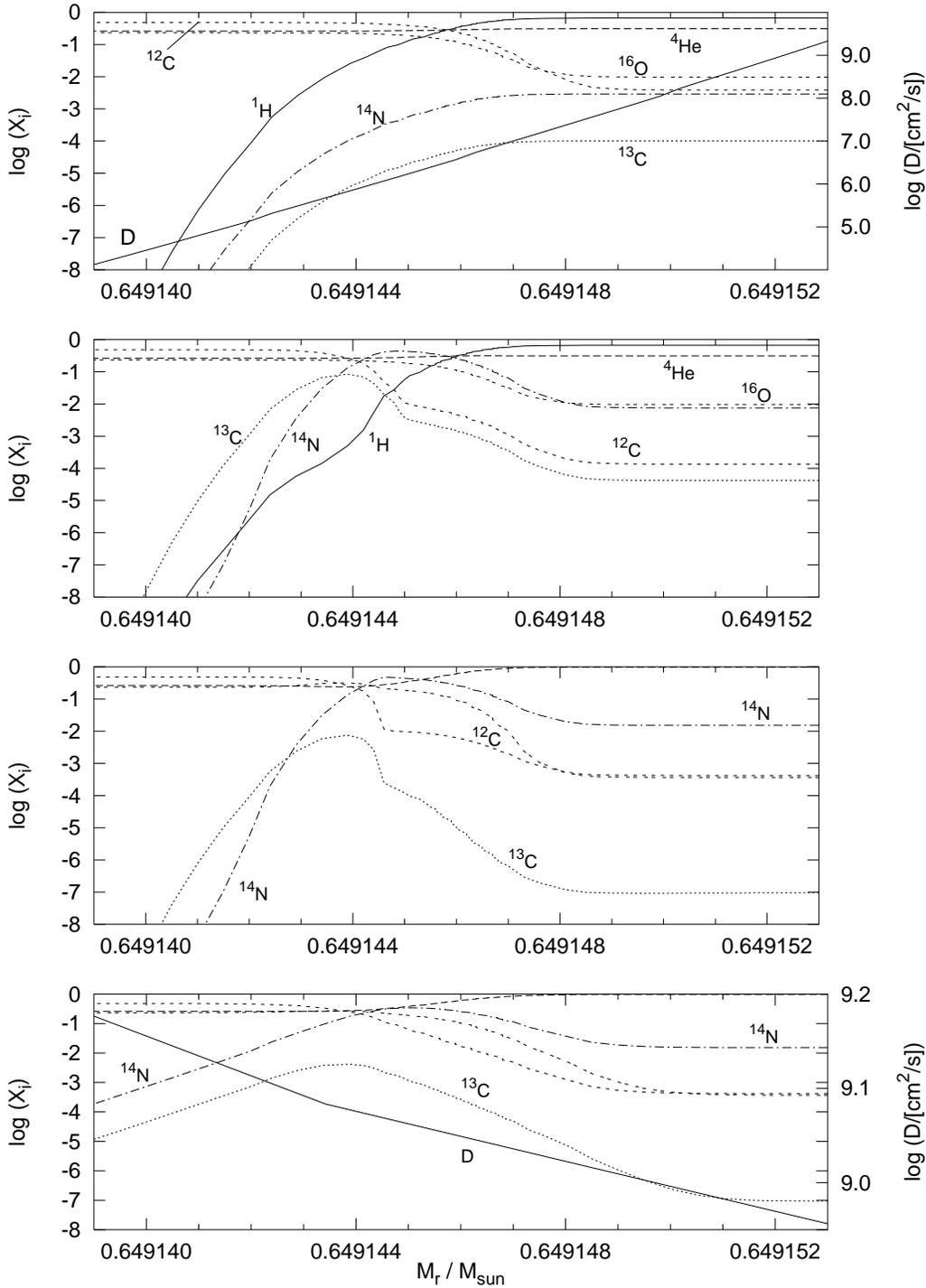} \hspace{0.5cm}
  \parbox[b]{3.5cm}{
    \caption[\cdr\ pocket]{ \label{C13I} 
      Development and destruction of the \cdr\ pocket after the eighth
      TP of the 3\msun\ sequence. The panels show the abundance
      profiles of H,\czw\ and other isotopes in a mass range that
      contains the contact region of envelope and core at the end of
      the TDUP.  The top panel shows a model after the end of the
      dredge-up episode (about $500 \jahre$ after the He-flash). The
      second panel shows a model about $1800 \jahre$ later when
      hydrogen burning has set in again. The third panel shows a model
      at the very end of the interpulse phase when \cdr\ has already
      been destroyed by reaction $\cdr(\alpha,n)\ose$. The last panel
      shows the profiles at the onset of the next thermal pulse. The
      solid lines in the top and bottom panel are the diffusion
      coefficient (right scale) of these models. In the top panel the
      diffusion coefficient results from the overshoot of the envelope
      convection while in the bottom panel it is the diffusion
      coefficient of the upper overshoot zone of the He-flash
      convection zone during the ninth pulse}.  }
\end{figure*}
\subsection{The formation of the \cdr\ pocket and of an adjacent \nvi\ pocket}
\label{cdr-main}

During the TDUP the hydrogen rich and convectively unstable envelope
has contact with the radiative carbon rich intershell region.  If the
boundary of the envelope convection zone is treated according to the
exponential diffusive overshoot method hydrogen diffuses into the
carbon rich layer below.  When the bottom of the envelope convection
zone has reached the deepest position by mass, a region forms were
protons from the envelope and \czw\ from the intershell region coexist
(\abb{C13I}, top panel). Note that the whole mass range shown in
\abb{C13I} is part of the overshoot region of the convective envelope.
The solid line in the top panel shows the exponentially declining
diffusion coefficient at the end of the dredge-up episode. In the
right half of the panel $D$ is large enough to cause a homogeneous
element mixture. The profiles have developed cumulatively over a
period of \mbox{$\approx 50 \jahre$} during which the mass coordinate
of the base of the envelope convection remains almost constant.  In
the left part the hydrogen abundance decreases as \czw\ increases.
Thus, the H/\czw\ number ratio decreases continuously from $\sim 750$
in the envelope to zero in the intershell region.

When the temperature increases during the succeeding evolution proton
captures transform \czw\ and H according to the continuously
decreasing H/\czw\ ratio into a \cdr\ pocket and a prominent \nvi\ 
pocket side by side.  The \cdr\ forms in the deeper layers where the
H-abundance is below $\simeq 5\%$.  Here the newly formed \cdr\ cannot
be further processed into \nvi\ because all protons are already
consumed. The \cdr\ abundance reaches a maximum mass fraction of
$0.09$ at $M_\mem{r}=0.649144\msun$.  Above this mass coordinate the
H/\czw\ ratio is larger and after the formation of some \cdr\ there
are still protons available which transform \cdr\ into \nvi.  The
\cdr/\nvi\ ratio changes continuously in this region and at any
position it is almost entirely determined by the initially present
H/\czw\ ratio.  Moreover, a \nvi\ pocket is a direct consequence of
the continuous variation of the H and \czw\ abundances from the core
to the envelope. Therefore, if the efficiency of mixing is decreasing
smoothly in the overshoot region a substantial \nvi\ pocket will
inevitably form.  The maximum abundance is given by the amount of
\czw\ in the intershell (see \kap{abundance-intershell-overshoot}). In
a narrow region (albeit larger than the region occupied by the \cdr\ 
pocket) \nvi\ becomes the most abundant isotope at a mass fraction of
$0.45$.  The bend of the profile, e.g.\ at $M_\mem{r}=0.649146\msun$
originates from the changing abundance ratios of \czw, H and \nvi\ 
which are relevant for the production of \cdr.

Note that at the time when proton captures start to form the \cdr\ and
\nvi\ pockets the radiative region \emph{above} the burning region is
well established. The isotopes made in this region cannot reach the
surface at this time. Instead they are processed under radiative
conditions \cite{straniero:95} and the products of this processing
will be engulfed by the next He-flash convection zone. Thus, the
production of \cdr\ after dredge-up with overshoot does not decrease
the observed \czw/\cdr\ ratio in giant stars as suspected by
Wallerstein \& Knapp \cite*{wallerstein:98}.

During the whole interpulse period the region displayed in \abb{C13I}
is heating and contracting (\abb{Rho-T_C13_max-TP8-M3eta1}). When the
temperature has reached about $T=10^8\kelv$, \cdr\ is destroyed
(\abb{C13_max-TP5-8}) by $\cdr(\alpha,n)\ose$ and neutrons are
released. Towards the end of the interpulse period (third panel
\abb{C13I}) most of the \cdr\ is burnt by $\alpha$ capture. The
remaining small fraction is engulfed by the outwards reaching He-flash
convection zone of the ninth pulse.  This is displayed in the bottom
panel of \abb{C13I}. The straight solid line shows the diffusion
coefficient in the overshoot region at the top of this convection
zone. Although the absolute value of $D$ is larger than in the top
panel, the impact on the abundance appears to be smaller. This effect
is caused by the smaller velocity of the convective boundary of the
envelope during the dredge-up compared to that of the boundary of the
He-flash convection zone.

\abb{C13_max-TP5-8} shows that after more advanced pulses (like the
eighth in this case), the \cdr\ formed at the hydrogen -- carbon
interface does indeed burn almost completely under radiative
conditions \cite{straniero:95}.  The \cdr\ which formed after earlier
pulses (like the fifth in this case) burns only partly under radiative
conditions. After early thermal pulses a certain fraction of the \cdr\ 
will be mixed into the \pdcz\ of the succeeding TP where it is
processed under convective conditions in the intershell.  In
\abb{Rho-T_C13_max-TP8-M3eta1} the density and temperature at the
position of the \cdr\ pocket are given for the case of the eighth TP.
This shows the conditions under which the neutron capture operates.

\begin{figure}[tbhp] 
  \epsfxsize=8.8cm \epsfbox{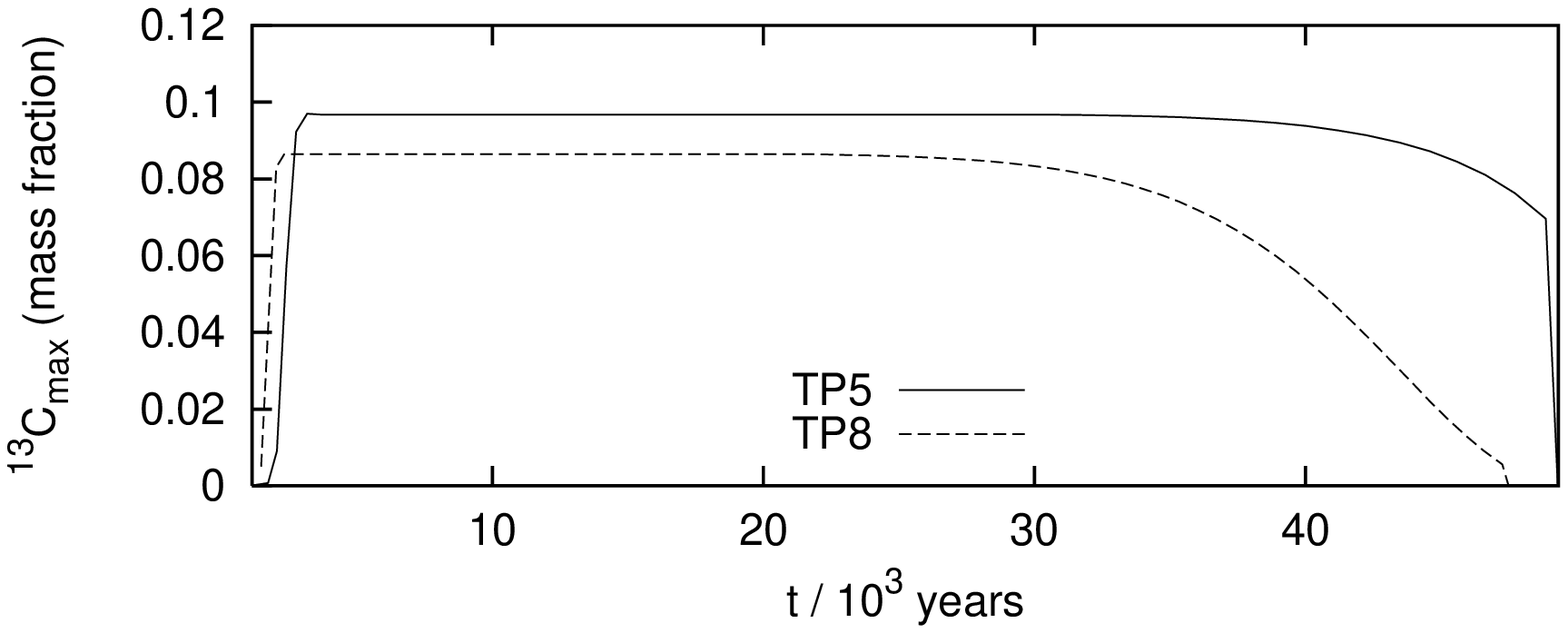}
\caption[Evolution of \cdr\ pocket]{ \label{C13_max-TP5-8} 
  The maximum \cdr\ abundance of the pocket after the fifth and eighth
  pulse of the $3\msun$ sequence displayed over one pulse cycle
  respectively. The zero time is set to the flash peak respectively.
  }
\end{figure} 
\begin{figure}[tbhp] 
  \epsfxsize=8.8cm \epsfbox{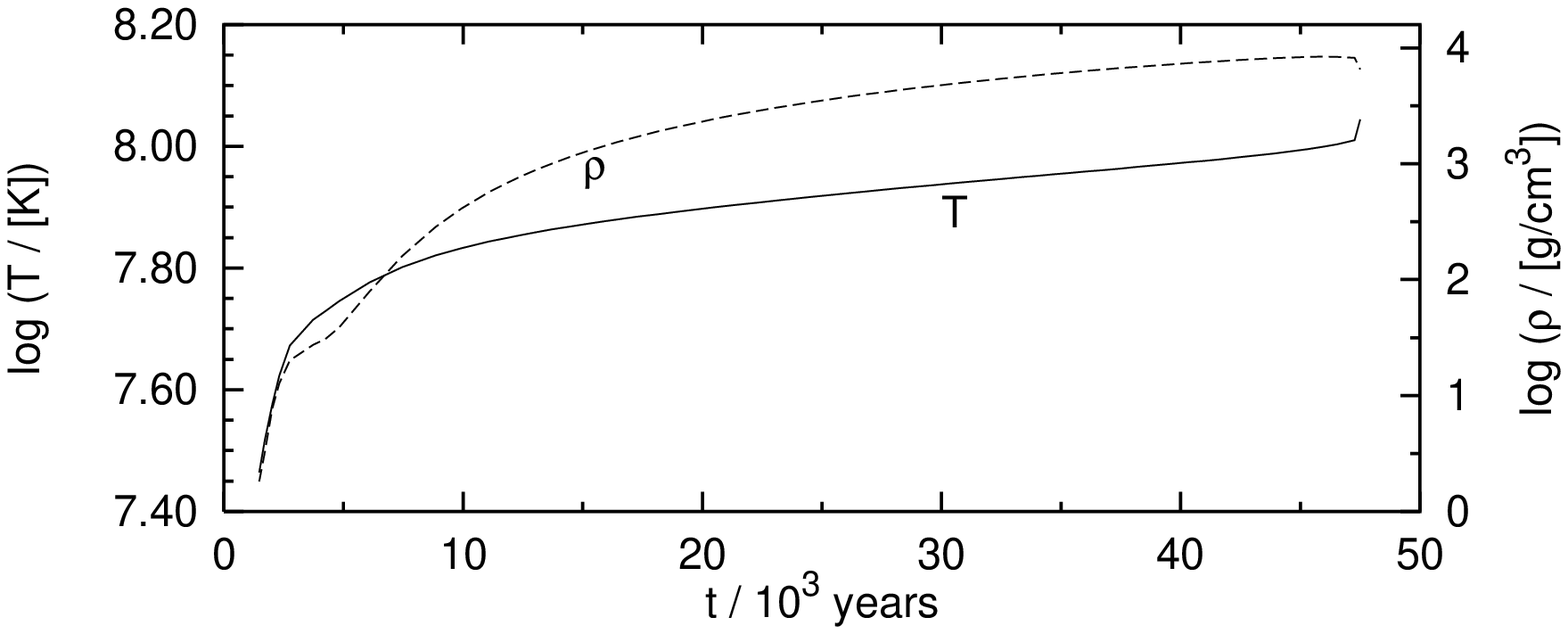}
\caption[Temperature and density in \cdr\ pocket]{ \label{Rho-T_C13_max-TP8-M3eta1} 
  The temperature and density at the location of the maximum \cdr\ 
  abundance for the eighth TP corresponding to \abb{C13_max-TP5-8}.  }
\end{figure} 
With our overshoot description and assumption about the efficiency
($f=0.016$) the layer in which the \cdr\ pocket forms is very thin
(see \abb{C13I}). The total mass of \cdr\ contained in the pocket
comprises only $\approx\nat{2\dots4}{-7}\msun$.  Estimates of the
\spr\ element distribution from the \cdr\ neutron source demand that
the pocket contains $\approx\nat{1\dots2}{-5}\msun$ in order to model
the main \spr\ component \cite{straniero:95}. However, the numbers are
not comparable at face value because the result of Straniero \etal\ is
based on models without overshoot.  Their hydrogen profile in the
carbon rich region is not the same as our overshoot hydrogen profile.
Also, the contribution and modification of the $s$-process element
distribution from the $\nezw(\alpha,n)\mgfu$ neutron source reaction
during the high-temperature phase of the flash should be different in
our models (see \kap{struct-intershell-overshoot} and
\ref{sec:3msun-models}).  Finally, the additional \nvi\ pocket which
forms due to the overshoot model must be considered. On one hand \nvi\ 
is known to be an important neutron poison because of its large
$(n,p)$ cross section. On the other hand a large fraction of the
neutrons lost by this reaction may be reproduced by
$\czw(p,n)\ndr(\beta^+)\cdr(\alpha,n)\ose$.  Moreover the additional
amount of \nvi\ to be ingested into the next He-flash convection zone
will also be converted into additional \nezw. All those processes will
affect the $s$-process nucleosynthesis, which cannot be reduced to the
mere amount of \cdr\ produced.

Preliminary tests with different overshoot efficiencies have shown
that the shape of the abundance profiles in the region of the \cdr\ 
pocket is scaled with respect to the mass coordinate but conserved
with respect to the abundance ratios. Intervals with certain mass
ratios of, e.g. \czw\ and hydrogen are just larger with a larger $f$.
This means that a dedicated study of the $s$-process nucleosynthesis
with a \cdr\ pocket according to our overshoot description should find
that within a certain range of efficiencies the f-value does only
determine the total amount of $s$-process elements in the dredged-up
material but not its distribution. It should be investigated whether
the functional form of an exponential velocity decay can reproduce a
$s$-process distribution in compliance with the solar main component.

Apart from overshoot the process of rotationally induced mixing is a
promising mechanism for the origin of the \cdr\ pocket
\cite{langer:99}. It is surprising that the overall amount of \cdr\ 
found in the pocket due to rotation is almost identical to the amount
found here with overshoot (a few $10^{-7}\msun$).

\section{The role of overshoot at the boundaries of the He-flash convection
  zone}
\label{sec:overshoot-at-pdczb}

During the He-flash the intershell region becomes convectively
unstable due to the huge energy generation of the He-burning shell.
Overshoot at the bottom of this He-flash convection zone leads to a
deeper mixing from the region below the He-shell (the C/O core) into
the intershell. This process could be called the fourth dredge-up
\cite{iben:99} or intershell dredge-up.  Both the abundance in the
intershell and also the structural parameters in the convection zone
are affected by overshoot at the base of the He-flash convection zone.
In particular, the change of the intershell abundance and the larger
energy generation during the TP leads to very efficient TDUP
\cite{herwig:98c}. If these overshoot AGB models are used as starting
models for the post-AGB evolution the long-standing discrepancy
between observed surface abundances of H-deficient post-AGB stars and
stellar models of this phase can be resolved \cite{herwig:99c}.
\begin{figure}[htp]
  \epsfxsize=8.8cm \epsfbox{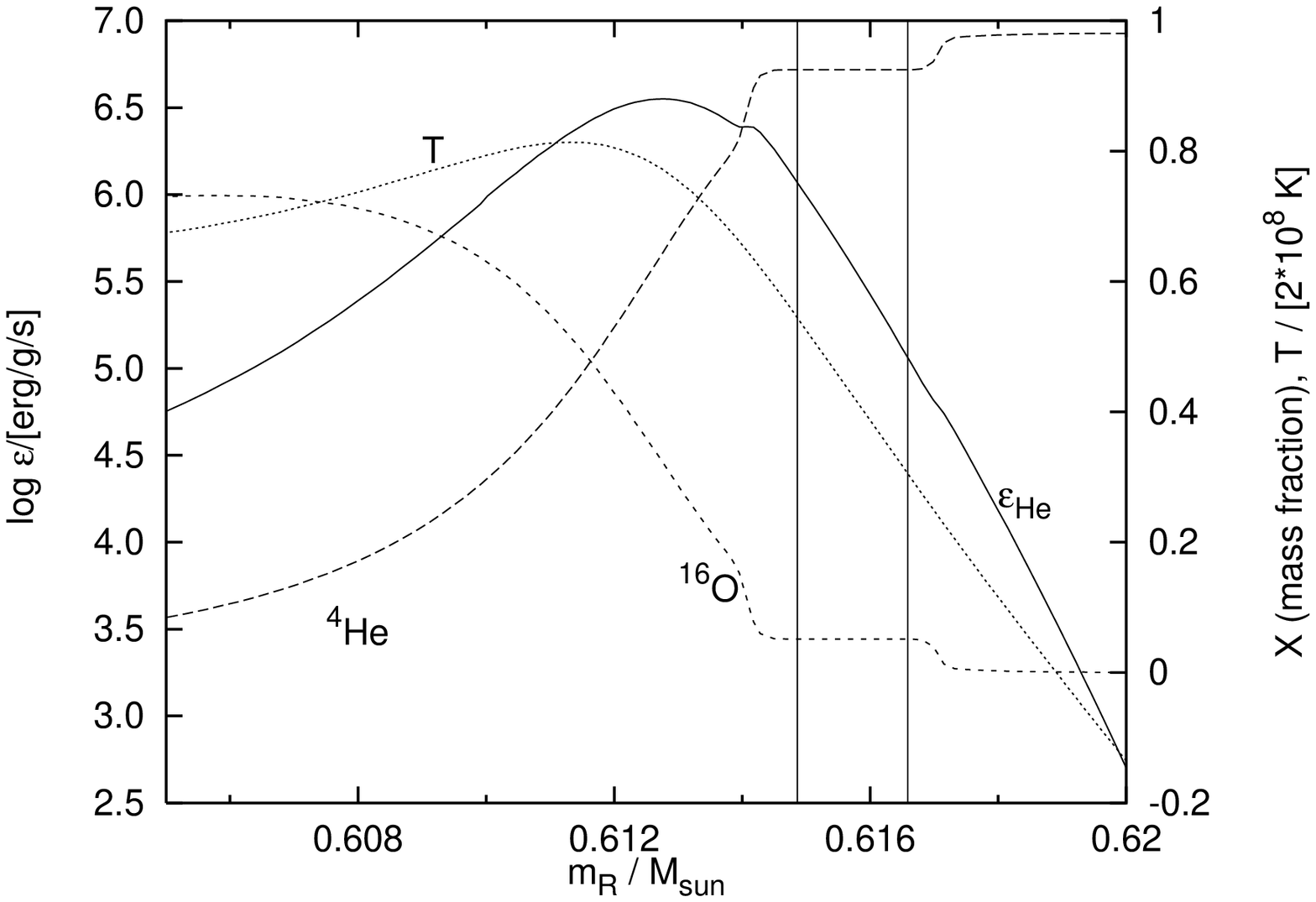}
\caption[]{ \label{fig:PDCZ8693OV}
  Onset of He-flash convective instability in the intershell region at
  the first TP of 3\msun\ sequence with overshoot. The two vertical
  lines show the upper (right) and lower (left) boundary of
  convection.}
\end{figure}
\subsection{Intershell structure during the flash}
\label{struct-intershell-overshoot}

During the onset of the He-flash, when helium-burning increases
rapidly, the layers just above the position of maximum energy
generation and temperature become convectively unstable and the
pulse-driven convection zone develops (\abb{fig:PDCZ8693OV}).  In this
situation overshoot at the bottom of the PDCZ supplies additional
fresh helium into the underlying nuclear burning region and supports
the nuclear runaway.  The situation is somewhat similar to hot-bottom
burning were the hydrogen burning shell obtains additional nuclear
fuel from the deep envelope convection and accordingly increases its
energy production.  Similarly the extra mixing below the PDCZ leads to
a more violent He-flash with larger peak He-burning luminosities
(\abb{fig:Mc-LHe}) and larger mass $\Delta M_\mathrm{PDCZ}$ of the
convectively unstable region (\abb{fig:mass-all}d).  The stronger
He-flashes are responsible for the structural and abundance
consequences discussed below.  The larger extent of the He-flash
convection zone has immediate consequences, e.g. for the \spr\ 
nucleosynthesis: Heavy elements produced during the interpulse phase
under radiative conditions are diluted in the convective intershell
during the pulse before they are dredged-up into the envelope. For a
larger He-flash convection zone the dilution effect will be larger.

Apparently there is no noticeable effect of overshoot at the top of
the PDCZ. For the metallicity and overshoot efficiency considered here
the minimum mass layer remaining between the H-rich envelope and the
top of the PDCZ at its largest extension is independent on overshoot
as can be seen from \abb{fig:mass-all}c.

The sequence without overshoot does not show any TDUP, and accordingly
the whole intershell with the He-burning, and the H-burning shell at
each boundary respectively is shifted outward by mass. For the 3\msun\ 
sequence with overshoot dredge-up sets in at the third TP and the
effective core mass growth ($\Delta M_\mathrm{H} - \Delta
M_\mathrm{DUP}$) is slowed down (\abb{fig:mass-all}a,b).  At the last
computed TP of the 3\msun\ sequence the intershell no longer evolves
outward with mass. During this stationary shell burning the nuclear
fuel for both the hydrogen and helium shell is transported downward by
dredge-up. Over the pulse cycle the dredged-down hydrogen is first
converted into helium which is then further mixed down by the He-flash
convection zone. Finally the burning products are exchanged by
dredge-up with fresh envelope material. In this situation dredge-up is
not only important to mix processed material up but also to mix fresh
material down and feed the stationary shells.

For the 4\msun\ sequence the core mass actually decreases from pulse
to pulse because $\lambda>1$ (\tab{4Meta1f}).  In that case the
nuclear burning shells are shifted inward with respect to mass.
Obviously this situation can only last because fresh material for the
nucleosynthesis is transported downward by dredge-up.

At the fifth TP (3\msun) the core masses of the sequence with and
without overshoot have not yet diverged too much due to the dredge-up
difference between the two cases. However, the differences due to
overshoot are already well established.  Not only is the mass-wise
extent of the PDCZ larger with overshoot (\abb{fig:mass-all}d) but in
addition the convective instability is longer
(\abb{fig:TP5-STRUC-all}a).  The He-flash convection zone develops and
disappears faster with overshoot. The overall duration of convective
instability in the intershell is about $100 \jahre$ with overshoot and
more than $150 \jahre$ for the case without overshoot.

With overshoot the temperature at the bottom of the He-flash
convection zone is \emph{larger} while at the top of the convectively
unstable zone the temperature is \emph{smaller}
(\abb{fig:TP5-STRUC-all}b).  Although the overall duration of the
convective instability is larger without overshoot it is important to
note that the high-$T$ phase is much longer with overshoot. The
duration of the high-T phase and the temperature reached during this
phase is important for the analysis of the nucleosynthesis during the
He-flash, in particular of the \nezw\ neutron source
($\nezw(\alpha,\n)\mgfu$).  Without overshoot the maximum temperature
reached at the bottom of the He-flash convection zone during the fifth
TP is $\nat{2.44}{8}\kelv$.  With overshoot the temperature at the
bottom of the He-flash convection zone exceeds $\nat{2.5}{8}\kelv$ for
$21.8 \jahre$ while $T>\nat{2.7}{8}\kelv$ for $9.3 \jahre$. Over the
following TPs the temperature in the He-flash convection will increase
steadily. However, the general trend described above will remain
preserved.  Therefore, the exact mechanism of the \spr\ and also other
aspects of the nucleosynthesis will be affected not only by the
different intershell abundances due to overshoot but also because of
different temperatures and time scales in the intershell during the
He-flash.

\abb{fig:TP5-STRUC-all}c displays the geometric evolution of the
He-flash convection zone together with the position of the maximum
energy generation by helium burning. The fact that the temperature at
the top of the He-flash convection zone is lower with overshoot is
closely related to the greater expansion of the intershell in this
case. With overshoot the largest geometric extent of the He-flash
convection zone is about $1.4$ times the distance found without
overshoot.

\begin{figure}[htp]
  \epsfxsize=8.8cm \epsfbox{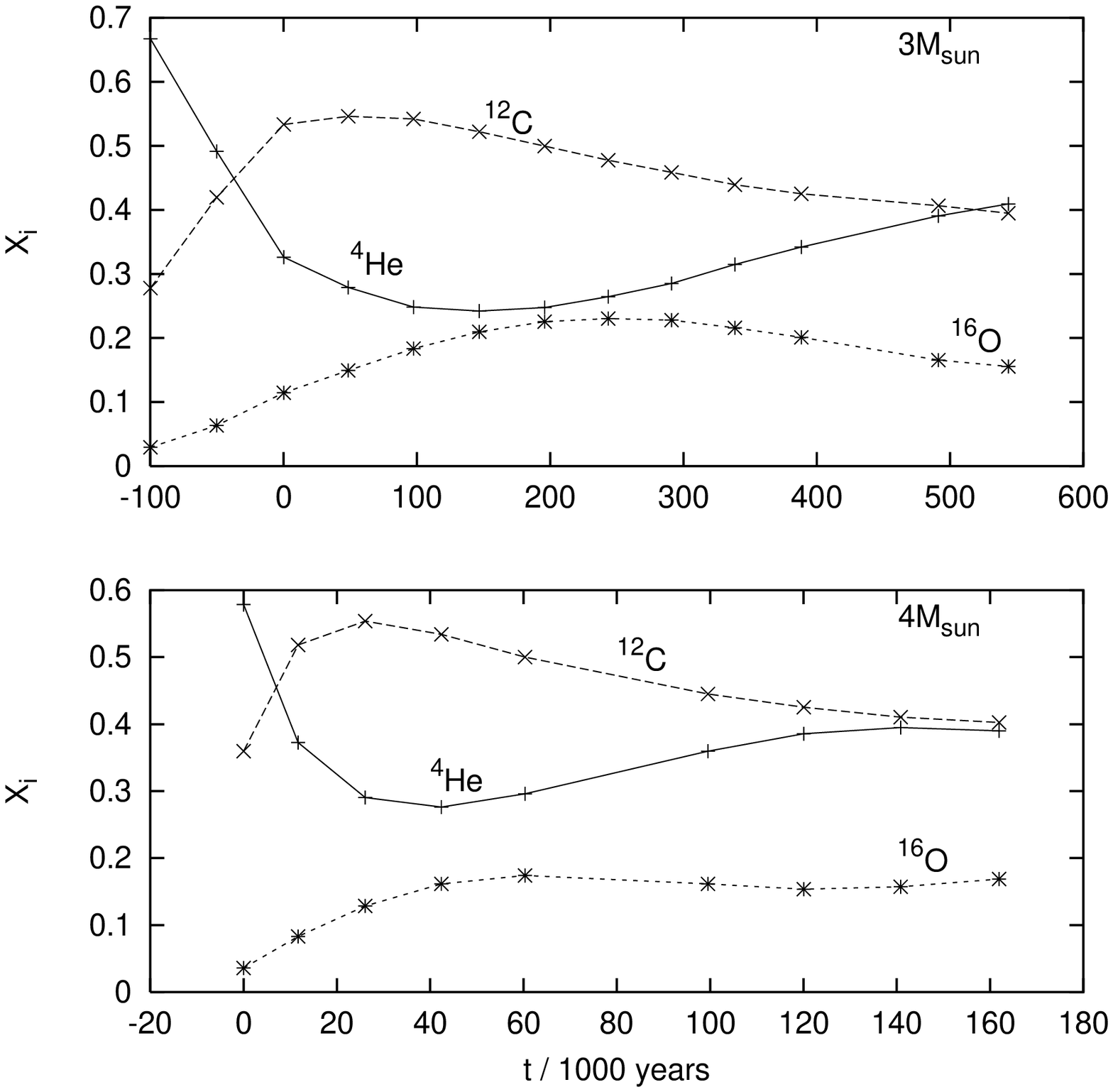}
\caption[Evolution of intershell abundances]{ \label{ZSH-t}
  Time evolution of abundances (mass fraction) in the upper part of
  the intershell shortly after each TP of the $3\msun$ and $4\msun$
  model sequence with overshoot (compare with Boothroyd \& Sackmann
  \cite*{boothroyd:88c}, Fig.\,9).  Each mark indicates one thermal
  pulse starting at the first TP.  The zero point for the $3\msun$
  sequence has been set at the third TP where the TDUP stars. }
\end{figure}
\begin{figure}[th]
  \epsfxsize=8.8cm \epsfbox{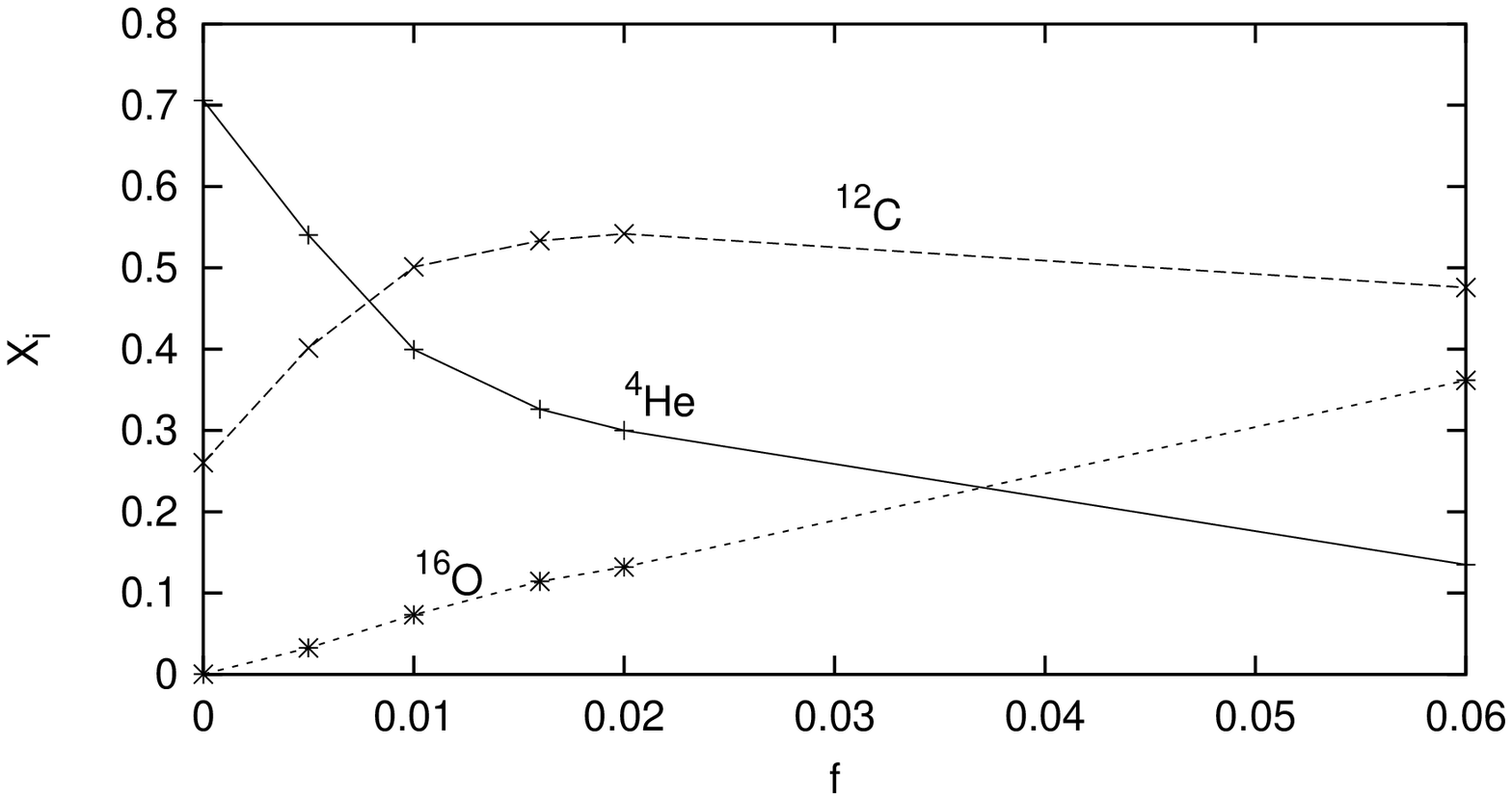}
\caption[]{
\label{f-Xi-TP3} 
Intershell abundances for different overshoot efficiencies after the
third TP. Test model sequences have been calculated from the same
starting model before the first TP with different values of $f$.  The
abundances for the sequence with $f=0$ are computed without
overshoot.}
\end{figure}

\subsection{The abundances in the intershell region}
\label{abundance-intershell-overshoot}
Models with overshoot have larger intershell mass fractions of carbon
and oxygen at the expense of helium. In \abb{ZSH-t} the variation of
the intershell abundances from pulse to pulse is shown. These models
with overshoot show qualitatively the same dependence on the pulse
number like models without overshoot
\cite{schoenberner:79,boothroyd:88} but the quantitative abundances
are very different. During the first TPs the \czw\ and \ose\ abundance
increases strongly at the expense of \hevi. After about six TPs the
\hevi\ abundance in the intershell reaches a minimum while \czw\ and
\ose\ go through a maximum. After about five to ten additional TPs all
abundances are leveling at values which are similar to those found
after the second or third TP.  Quantitatively, overshoot with the
efficiency $f=0.016$ increases the amount of carbon by about a factor
of two and the amount of oxygen by a factor of 10 to 20.

Test calculations in order to study the dependence of the intershell
abundance on the efficiency of overshoot reveal that there is a
relation between the overhoot parameter and the intershell abundance:
larger overshoot leads to larger carbon and oxygen abundances and
consequently smaller helium abundance. We have followed the evolution
from the same starting model before the first TP over three TPs with
different values of $f$. In \abb{f-Xi-TP3} we display the abundance in
the upper part of the intershell after the third TP. For $f=0$ the
abundances are the same as in Sch\"onberner \cite*{schoenberner:79}.
With larger and larger overshoot efficiency, the amount of oxygen
increases almost linearly.

This correlation leads to a quite stringent constraint for the
possible range for the overshoot efficiency $f$ at the bottom of
He-flash convection zone. We compare our intershell abundances with
the surface abundances of the [WC]-type central stars of planetary
nebulae \cite{koesterke:97b,hamann:97,demarco:97} and the PG\,1159
stars \cite{dreizler:98,werner:98}.  The spectroscopic abundance
analysis shows that these stars are very carbon rich and also oxygen
rich. Typically, one finds (He,C,O)=(0.50,0.33,0.17) (mass fractions)
for PG\,1159 stars.  Post-AGB stars become hydrogen-deficient because
the intershell material appears at the surface in the aftermath of a
very late TP \cite{herwig:99c}.  Thus, the observed surface abundances
of H-deficient post-AGB stars must be interpreted as the intershell
abundances of the progenitor AGB stars \cite{herwig:99d}.  Models
without overshoot never show more than $2\%$ of oxygen in the
intershell and are thus unable to reproduce the large mass fraction of
oxygen observed. This is a strong indication of the presence of at
least some overshoot at the bottom of the He-flash convection zone.
However, practically no H-deficient post-AGB stars are known to show
more than about $20\%$ of oxygen\footnote{The exception, H1504+65, is
  free of H \emph{and} He and has C $\simeq$ O.}.

The oxygen abundances shown in \abb{f-Xi-TP3} are an estimate of the
surface oxygen abundance of a H-deficient post-AGB model evolved from
the respective AGB model because the oxygen abundances after the third
TP is already similar to the later TPs (\abb{ZSH-t}).  Any overshoot
parameter much larger than $f=0.03$ would lead to H-deficient post-AGB
models with too large an oxygen abundance. Therefore, observations -
together with the theoretical understanding of the evolutionary origin
of H-deficient post-AGB stars - do constrain the overshoot efficiency
at the bottom of the He-flash convection zone to a narrow range of
$0.01 \apleq f \apleq 0.03$.

\subsection{The third dredge-up and its dependence on overshoot at the 
  bottom of the He-flash convection zone}
\label{sec:oversh-dredge}
One of the most important properties of TP-AGB models with overhoot is
the efficient TDUP, which solves the problem of previous models to
account for the typically low luminosities of carbon stars. In
\tab{3Meta1f} and \ref{4Meta1f} the amount of dredged-up intershell
mass and the related values for the dredge-up efficiency $\lambda =
\Delta M_\mathrm{\lambda} / \Delta M_\mathrm{H}$\footnote{$\lambda$ =
  mass dredged-up after TP / growth of intershell by hydrogen burning
  during the preceding interpulse phase.} are given for the 3\msun\ 
and 4\msun\ sequence with overshoot, and these results will be further
discussed in \kap{sec:3msun-models}.

Here, we consider the dependence of $\lambda$ on the efficiency of
overshoot. In particular it is important to clarify the different
roles of overshoot below the bottom of the He-flash convection zone
and below the bottom of the envelope convection zone (see also
\kap{sec:tdup}). For this purpose we have made a numerical experiment
which involved the computation of one TP cycle with different values
for the overshoot efficiency parameter $f$. The computations were
started immediately after the He-flash convection zone of the eighth
TP had disappeared. Thus, the dredge-up episode following this TP has
been computed with different $f$ values at the envelope convection
zone only, while during the preceding TP $f$ was the same for all
convective boundaries. The resulting values for $\lambda$ are shown in
\abb{fig:f-lambda} as open symbols and demonstrate the dependence of
the third dredge-up on the overshoot efficiency below the convective
envelope. Within the considered range of $f$-values practically no
dependence of $\lambda$ on the overshoot efficiency exists. Only
without any overshoot ($f=0$) does dredge-up not occur for this TP.
Even with $f$ as small as $0.004$ dredge-up is found with practically
the same efficiency than with larger $f$ (see \kap{sec:tdup}). Note
that the value of $\lambda \simeq 0.7$ displayed for the open symbols
in \abb{fig:f-lambda} is just the value found for the eighth TP of the
3\msun\ overshoot sequence (\tab{3Meta1f}).  If another TP had been
chosen for this numerical experiment $\lambda$ would have had the
correspondingly larger or smaller value.  \abb{fig:f-lambda} does not
mean that models with overshoot applied only to the envelope
convection always have $\lambda \simeq 0.7$.

The test model sequences have then been evolved beyond the dredge-up
episode after the ninth TP. The respective $f$-values have now been
applied to the convective boundaries of the He-flash convection zone
as well. The resulting $\lambda$-values are represented by filled
symbols in \abb{fig:f-lambda} and show the dependence of the dredge-up
on the overshoot efficiency at the bottom of the He-flash convection
zone.  The full symbol at $f=0.016$ shows $\lambda \simeq 0.8$, the
value given in \tab{3Meta1f} for the ninth TP.  $f$ and $\lambda$ are
correlated: larger overshoot leads to larger dredge-up. The relation
defined by the full symbols is specific to the chosen TP of this
sequence.  However, the trend is generally valid.  A combination of
effects is responsible for this correlation. With intershell overshoot
the peak He-burning luminosities are substantially larger
(\abb{fig:Mc-LHe}). At the 14$^\mem{th}$ TP of the overshoot sequence
($\mh = 0.658\msun$) the He-flash peak luminosity is $\lhe =
\nat{7.9}{7} \lsun$.  At the 14$^\mem{th}$ TP of the sequence without
overshoot ($\mh = 0.692\msun$) the peak luminosity is $\lhe =
\nat{2.4}{7} \lsun$.  Boothroyd \& Sackmann \cite*{boothroyd:88c}
found for a 3\msun, Z=0.02 model sequence $\lhe \simeq \nat{2}{7}
\lsun$ after about 20 TPs at a core mass $\mh \simeq 0.65\msun$. These
models were computed without overshoot during the pre-AGB evolution
and therefore have a lower core mass at the first TP $M_\mem{H1}
\simeq 0.55\msun$.

Greater He-flash strength leads to a greater expansion and cooling of
the upper layers of the intershell (\abb{fig:TP5-STRUC-all}) and
favors the occurrence of the TDUP ($\gradrad \propto T^{-4}$).  In
\abb{f-ABUND-TP8} one can see that the modification of the intershell
abundance during only one TP is considerable. The smaller He abundance
with larger $f$ leads to a larger opacity $\kappa$ which favors the
dredge-up also because $\gradrad \propto \kappa$.

Thus the overshoot at the base of the He-flash convection zone
strongly affects TDUP.  In the previous section we have shown that $f$
is constrained by the observational properties of H-deficient post-AGB
stars. Therefore we conclude that the upper limit of the efficiency of
TDUP can be constrained from observational properties of post-AGB
stars.

\begin{table*}[hbtp]
\caption[\Dup\ properties of the $3\msun$ sequence]{\Dup\ properties of the thermal pulses
of the $3\msun$ sequence with
$f=0.016$. The table gives the TP number, 
age (zero point set at first TP which shows \dup),
peak luminosity of the helium burning shell $L_\mem{He} / \lsun$, 
amount of dredged-up hydrogen free material $\Delta M_\lambda /
\msun$, core growth since last TP $\Delta M_\mem{H} / \msun$,
\dup\ parameter $\lambda$ (see text), mass of hydrogen free core
$M_\mem{H}$,  mass coordinate of top of \pdcz\ $M_\mem{Pt}$,
mass coordinate of  bottom of \pdcz\ $M_\mem{Pb}$ and isotopic and
elemental abundance ratios at the surface immediately after the TP.
} \label{3Meta1f}
\begin{center} 
\begin{tabular}{rrlllllllrrrr} 
\hline \noalign{\medskip}
TP  &$\frac{\mem{age}}{\jahre}$
&$\frac{L_\mem{He}}{10^6\lsun}$&$\frac{\Delta
M_\lambda}{10^{-3}}$&$\frac{\Delta
M_\mem{H}}{10^{-3}}$&$\lambda$&$\frac{M_\mem{H}}{\msun}$&$\frac{M_\mem{Pt}}{\msun}$&$\frac{M_\mem{Pb}}{\msun}$& 
$^{12/13}\mathrm{C}$ &C/O &$^{16/17}\mathrm{O}$ &$^{16/18}\mathrm{O}$ \\ 
\noalign{\medskip}
\hline      
$1$&$ -100079$&$ 0.46$&$ -   $&$-  $&$ -   $&$ 0.63087$&$
0.62969$&$   0.60598$ &$20.4 $&$0.29 $&$194.7 $&$606.7 $\\
$2$&$ -50212 $&$ 1.17$&$ -   $&$2.0$&$ -   $&$ 0.63288$&$
0.63213$&$   0.60795$ &$'' $&$ ''$&$ ''$&$'' $\\
$3$&$ 1      $&$ 2.13$&$ 0.03$&$3.1$&$ -   $&$ 0.63593$&$
0.63547$&$   0.61003$ &$'' $&$ ''$&$ ''$&$'' $\\
$4$&$ 48519  $&$ 2.63$&$ 0.4$&$ 3.7$&$ 0.10$&$ 0.63962$&$
0.63920$&$   0.61377$ &$20.5 $&$0.30 $&$194.8 $&$607.0 $\\
$5$&$ 97563  $&$ 3.18$&$ 1.0$&$ 4.3$&$ 0.23$&$ 0.64352$&$
0.64315$&$   0.61819$ &$22.3 $&$0.32 $&$196.1 $&$610.9 $\\
$6$&$ 146883 $&$ 4.63$&$ 2.1$&$ 4.7$&$ 0.44$&$ 0.64726$&$
0.64694$&$   0.62267$ &$26.9 $&$0.38 $&$199.8 $&$622.7 $\\
$7$&$ 195842 $&$ 6.71$&$ 2.7$&$ 5.1$&$ 0.53$&$ 0.65030$&$
0.65002$&$   0.62688$ &$33.0 $&$0.45 $&$205.5 $&$640.4 $\\
$8$&$ 243582 $&$ 8.80$&$ 3.7$&$ 5.3$&$ 0.71$&$ 0.65289$&$
0.65265$&$   0.63048$ &$41.4 $&$0.54 $&$213.9 $&$666.6 $\\
$9$&$ 291084 $&$ 13.9$&$ 4.6$&$ 5.7$&$ 0.82$&$ 0.65482$&$
0.65461$&$   0.63328$ &$51.8 $&$0.64 $&$224.5 $&$699.7 $\\
$10$&$ 338588 $&$ 19.0$&$ 5.4$&$ 6.0$&$ 0.91$&$ 0.65616$&$
0.65598$&$   0.63542$ &$63.8 $&$0.74 $&$236.6 $&$737.4 $\\
$11$&$ 388433 $&$ 29.4$&$ 6.1$&$ 6.4$&$ 0.98$&$ 0.65717$&$
0.65702$&$   0.63701$ &$77.7 $&$0.86 $&$250.0 $&$779.1 $\\
$12$&$ 438979 $&$ 42.9$&$ 7.0$&$ 6.8$&$ 1.02$&$ 0.65773$&$
0.65759$&$   0.63805$ &$93.0 $&$0.97 $&$263.7 $&$822.0 $\\
$13$&$ 491151 $&$ 59.9$&$ 7.5$&$ 7.3$&$ 1.04$&$ 0.65804$&$
 0.65804$&$  0.63871$ &$109.5 $&$1.08 $&$277.5 $&$864.9 $\\
\hline
\end{tabular}
\end{center}
\end{table*}
\begin{table*}[htpb]
\caption[\Dup\ properties of the $4\msun$ sequence]{\Dup\ properties of the thermal pulses
of the $4\msun$ sequence with
$f=0.016$. For explanation of the displayed quantities see \tab{3Meta1f}. 
} \label{4Meta1f}
\begin{center} 
\begin{tabular}{rrlllllllrrrr} 
\hline \noalign{\medskip}
TP  &$\frac{\mem{age}}{\jahre}$ &$\frac{L_\mem{He}}{10^6\lsun}$&$\frac{\Delta M_\lambda}{10^{-3}}$&$\frac{\Delta M_\mem{H}}{10^{-3}}$&$\lambda$&$\frac{M_\mem{H}}{\msun}$&$\frac{M_\mem{Pt}}{\msun}$&$\frac{M_\mem{Pb}}{\msun}$ & 
$^{12/13}\mathrm{C}$ &C/O &$^{16/17}\mathrm{O}$ &$^{16/18}\mathrm{O}$ \\ 
\noalign{\medskip}
\hline      
$1$&$ -80   $&$ 2.26$&$ 2.39 $&$ -    $&$-     $&$
0.78259$&$ 0.78239 $&$ 0.77172 $ &$19.8 $&$0.31 $&$274.3 $&$602.7 $\\
$2$&$ 11660 $&$ 3.70$&$ 3.12 $&$ 1.91 $&$ 1.63 $&$
0.78211$&$ 0.78192 $&$ 0.77096 $ &$22.4 $&$0.34 $&$275.1 $&$604.4 $\\
$3$&$ 25990 $&$ 6.21$&$ 3.94 $&$ 2.61 $&$ 1.51 $&$
0.78160$&$ 0.78144 $&$ 0.77017 $ &$27.7 $&$0.42 $&$277.6 $&$610.0 $\\
$4$&$ 42350 $&$ 11.5$&$ 4.79 $&$ 3.17 $&$ 1.51 $&$
0.78083$&$ 0.78070 $&$ 0.76945 $ &$34.9 $&$0.51 $&$282.7 $&$621.2 $\\
$5$&$ 60290 $&$ 21.0$&$ 5.41 $&$ 3.58 $&$ 1.51 $&$
0.77962$&$ 0.77953 $&$ 0.76865 $ &$43.5 $&$0.62 $&$290.6 $&$638.6 $\\
$6$&$ 79610 $&$ 38.1$&$ 5.63 $&$ 3.93 $&$ 1.43 $&$
0.77814$&$ 0.77806 $&$ 0.76766 $ &$52.8 $&$0.72 $&$300.4 $&$660.2 $\\
$7$&$ 99500 $&$ 59.8$&$ 5.90 $&$ 4.16 $&$ 1.42 $&$
0.77667$&$ 0.77661 $&$ 0.76656 $ &$62.0 $&$0.82 $&$310.4 $&$682.1 $\\
$8$&$ 120030$&$ 92.6$&$ 5.92 $&$ 4.45 $&$ 1.33 $&$
0.77522$&$ 0.77517 $&$ 0.76593 $ &$71.2 $&$0.91 $&$320.5 $&$704.4 $\\
$9$&$ 140790$&$ 123.3$&$0.61 $&$ 4.59 $&$ 1.33 $&$
0.77389$&$ 0.77384 $&$ 0.76472 $ &$80.1 $&$0.99 $&$330.2 $&$725.8 $\\
$10$&$ 161900$&$ 142.6$&$0.64 $&$ 4.74 $&$ 1.37 $&$
0.77253$&$ 0.77248 $&$ 0.76336 $ &$89.0 $&$1.06 $&$340.5 $&$748.6 $\\
$11$&$ 183480$&$ 195.5$&$0.65 $&$ 4.92 $&$ 1.32 $&$
0.77102$&$ 0.77128 $&$ 0.76187 $ &$98.3 $&$1.13 $&$352.4 $&$774.6 $\\
\hline
\end{tabular}
\end{center}
\end{table*}
\section{Surface properties}
\label{sec:3msun-models}
In the previous sections we have described the different mechanisms by
which overshoot influences the model properties. We will now focus on
the surface properties of the models with overshoot.

\paragraph{Stellar parameters}

The comparison of the core mass-luminosity relation of models with and
without efficient dredge-up shows clear differences \cite{herwig:98b}.
While models without dredge-up follow a linear relation when the
asymptotic regime has been reached, models with very efficient
dredge-up ($\lambda \simeq 1$) continue to increase in luminosity even
if the core mass no longer increases.  Here, the continuing radius
decrease leads -- according to simple homology relations
\cite{refsdal:70} -- to an increase in luminosities.  This effect is
most efficient over the earlier thermal pulses where the relative
radius decrease per TP cycle is larger than after many thermal pulses
when the core asymptotically resembles a white dwarf. The radius
effect is responsible for the sub-luminous phase (compared to the
luminosities expected from the core mass - luminosity relation) of the
first few TPs, which is well known from any TP-AGB model sequence.
However, the luminosity evolution continues to be strongly coupled to
the core radius evolution. This becomes apparent only if the core mass
is prevented from growing continuously in accordance with the radius
shrinkage, as in models with efficient dredge-up.  In addition, as
Marigo \etal\cite*{marigo:99a} have pointed out, possibly up to one
third of the luminosity increase observed by Herwig
\etal\cite*{herwig:98b} can be ascribed to the well known effect of
the molecular weight increase in the envelope as a result of dredge-up
of processed material.

The rather short-term variation of the stellar parameters during and
between the thermal pulses are affected as well. In general the
surface luminosity reacts to the TP in the deep interior by a sudden
($\Delta t \simeq \mathrm{thermal\ time scale} \simeq 100 \jahre$) and
drastic luminosity decline of about $40\%$ of the pre-TP luminosity.
It is followed (again on the thermal time scale) by an immediate
luminosity jump which forms the more or less pronounced \emph{TP
  surface luminosity peaks} (see \abb{fig:LSURF} at $t=0\jahre$). The
model sequence without overshoot shows only small TP-luminosity peaks
after the first TPs with peak luminosities lower than the quiescent
interpulse luminosity. Only after the third TP does the peak
luminosity exceed the interpulse luminosity and over the following TPs
the ratio of peak and interpulse luminosity grows.  The model sequence
with overshoot shows very large peaks already after the first TP with
a ratio of $\sim 1.6$ between the peak and the interpulse luminosity
which is gradually decreasing towards later TPs.  At the fifth TP
(\abb{fig:LSURF}) the surface luminosity peak is similar for the two
sequences. The actual local He-luminosity peak during a TP is followed
by a much smaller secondary peak. As this secondary maximum decays in
the intershell the surface luminosity does also decline to the second
surface luminosity minimum at $\sim 5000 \jahre$ in \abb{fig:LSURF}.
Another difference caused by overshoot is a faster recovery of the
H-burning shell after the TP (top panel \abb{fig:LSURF}). This is also
reflected by the evolution of the surface parameters.  For the model
sequence with overshoot it takes only $\simeq 7600 \jahre$ from the TP
until a $70\%$ level of the previous interpulse luminosity at the
surface has been regained while the sequence without overshoot
requires $\simeq 11800 \jahre$. This effect is even more pronounced at
later TPs, however it is then intermingled with the effect of
different core mass evolution due to the different efficiency of the
dredge-up.
\begin{figure}[t] 
  \epsfxsize=8.8cm \epsfbox{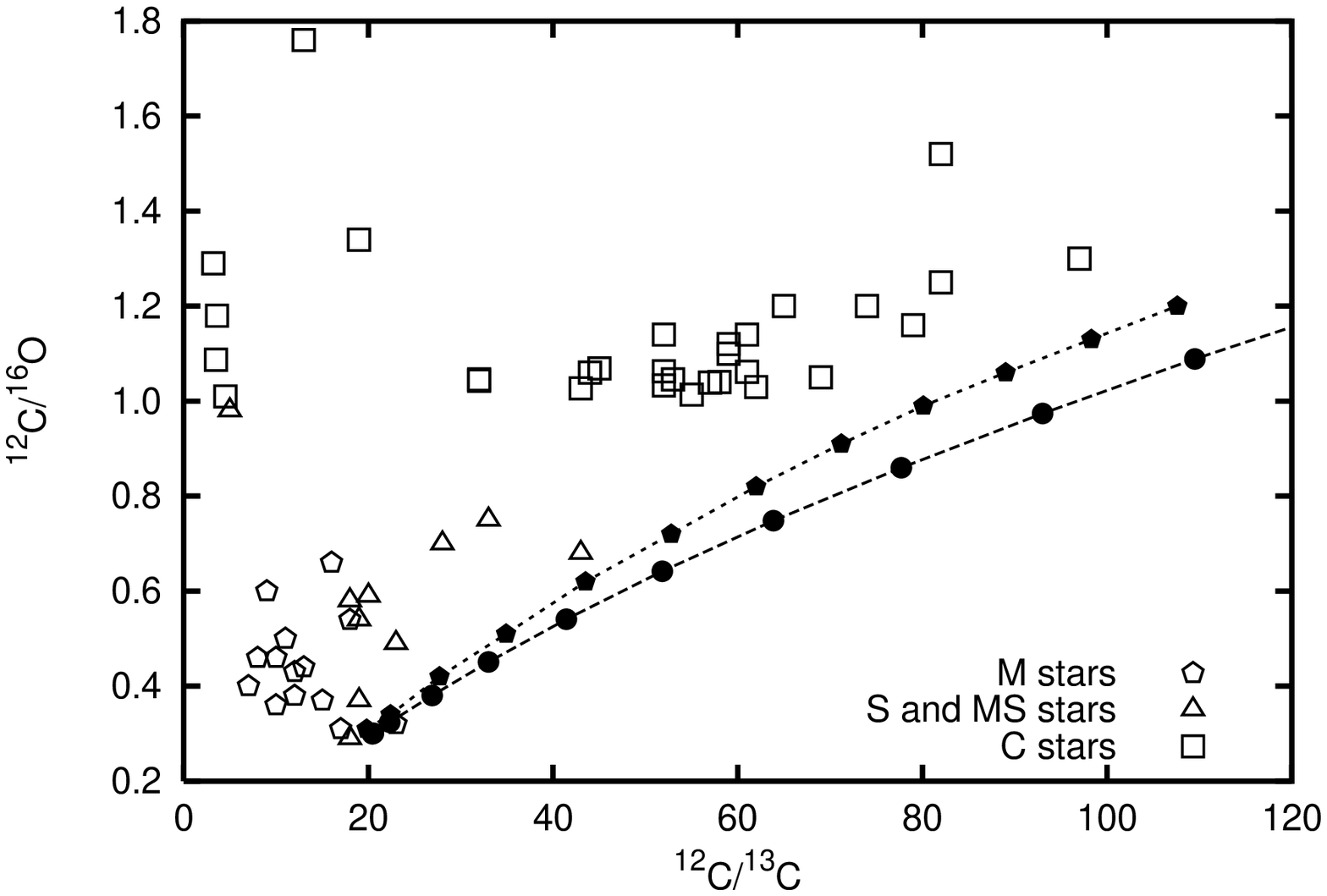}
\caption[]
{ \label{fig:C12_C13-CO} Reproduction of \czw/\ose\ versus \czw/\cdr\ 
  plot in Smith and Lambert (1990) (open symbols) together with the
  surface ratios of the $3\msun$ (filled circles) and $4\msun$ (filled
  pentagons) model sequence. See text for details.}
\end{figure} 
\nocite{smith:90}

\paragraph{Chemical evolution}

The chemical evolution of AGB models with overshoot is different
compared to models without overshoot due to four main effects:
\begin{enumerate}
\item The formation of a \cdr\ pocket in the upper part of the
  intershell region (\kap{cdr-main}). The consequences for the
  n-capture nucleosynthesis will be investigated elsewhere.
\item Substantially larger dredge-up, which leads to more efficient
  transport of processed material to the surface.
\item Higher temperature in He-flash convection zone during the TP
  (\abb{fig:TP5-STRUC-all}).
\item Depletion of helium and larger \ose\ and \czw\ abundance in the
  intershell due to deeper penetration of the bottom of the He-flash
  convection zone into the C/O-core.
\end{enumerate}
These effects are reflected by the chemical evolution at the stellar
surface (\tab{3Meta1f} and \ref{4Meta1f}).

\paragraph{C/O and carbon isotopic ratio}

Recurring and efficient \dup\ transports primary \czw\ into the
envelope after each TP. \cdr\ is not produced or destroyed in the
stars of the mass range discussed here. This leads to a steady
increase of the \czw/\cdr\ ratio at the surface. At the same time also
the C/O ratio increases from its initial value of $\sim 0.3$ and
exceeds unity after a number of TPs. Note that in calculations with
overshoot \ose\ is also mixed into the envelope at each \dup\ event.
Because of this additional \ose\ the C/O ratio grows more slowly with
overshoot than without overshoot if $\lambda$ where the same. However,
this effect is more than compensated for by two other effects: the
dredge-up itself is larger and the mass fraction of \czw\ and \ose\ of
the dredged-up intershell material is larger. This leads to a larger
increase of the C/O ratio for each TP. For example, as can be seen
from \tab{4Meta1f} the 4\msun\ sequence becomes C-rich at the tenth TP
contrary to the results of Forestini \& Charbonnel
\cite*{forestini:97} and Marigo \etal\cite*{marigo:96} who both find
that a 4\msun\ AGB star of solar metallicity does not become a C-star
at all. Note that these authors assume \dup\ to operate and that HBB
cannot prevent a 4\msun\ star from getting C-rich.

\abb{fig:C12_C13-CO} compares observational data of the chemical
composition of AGB stars with the model data of the sequences with
overshoot. The filled symbols in the lower left corner correspond to
the first TPs. The observational data shows properties of M- S- and
C-giant\footnote{M-giants:$\mem{C}/\mem{O}<0.8$, S-giants:
  $0.8<\mem{C}/\mem{O}<1.0$, C-giants:$\mem{C}/\mem{O}>1.0$.} stars
with different masses (most of them smaller than those of the model
sequences) which presumably also have experienced a different number
of TPs.\footnote{Giants without Tc have been excluded from
  \abb{fig:C12_C13-CO} because their abundance anomaly might be due to
  transfer of mantle material from a companion \cite{habing:96}.}
However, apart from giants with $\mem{C}/\mem{O} \approx 1$ and very
low $\czw/\cdr$ ratio (J-giants), a correlation between the two ratios
is apparent. With respect to the model data this correlation must be
interpreted as a time evolution. The model sequences do generally
follow the observed trend. However, the theoretical relation shows an
offset to the observations and also a somewhat smaller slope. In
addition the increment of the ratio from TP to TP is rather large and
depending on mass loss several more TP are likely for these model
sequences, leading to very large \czw/\cdr\ ratios. A somewhat smaller
overshoot efficiency would result in a lower dredge-up efficiency and
a smaller amount of dredged-up oxygen. While the first effect would
decrease the steps at which the model sequence passes along the
diagram, the latter would increase the slope of the theoretical
relation somewhat. As another detail, the initial abundance ratios
have a considerable influence on the theoretical prediction.  Most of
the displayed stars are probably of lower mass and therefore their
initial \czw/\cdr\ ratio before the first TP was closer to $\sim 10$
\cite{gilroy:89} than around $20$. This initial \czw/\cdr\ ratio
determines the the slope of the \czw/\cdr\ - C/O relation as well,
with a smaller initial ratio leading to a larger slope. Thus, models
with initial masses around or below 2\msun\ should show a larger slope
due to this effect and should show a better agreement with the
observations.

\paragraph{Oxygen isotopic ratios}

Observationally the oxygen isotopic ratios \ose/\osi\ and \ose/\oac\ 
of AGB stars extend to much larger values than predicted by
evolutionary models. Theoretical models without overshoot cannot
account for increasing \ose/\osi\ and \ose/\oac\ ratios because
dredge-up is only achieved with difficulty for the relevant masses.
Even if this problem is circumvented by assuming dredge-up to be
present then the dredged-up material leads only to a rather
inefficient dilution of \osi\ and \oac\ in the envelope because the
dredged-up material is depleted in these isotopes compared to the
envelope value. The \ose\ abundance in the intershell of models
without overshoot is similar to the envelope abundance and no
modification can be expected there.

Harris \etal\cite*{harris:85,harris:87} determined for \ose/\osi\ and
\ose/\oac\ values between $600$ and $3000$. It is also interesting to
note that the two ratios are correlated.  Moreover, both ratios seem
to be correlated with the neutron exposure associated with the
observed $s$-process abundances and also with the carbon abundance.
Both, an increasing carbon abundance and an increasing neutron
exposure are indicating more efficient or more frequent dredge-up
because $s$-process elements and carbon are not made in or at the
bottom of the envelope.  It is therefore likely that the enormous
oxygen isotopic ratios are related to the TDUP.

Forestini \& Charbonnel \cite*{forestini:97} have assumed dredge-up
and predict from an extrapolation of their models without overshoot an
increase of the \ose/\osi\ ratio of $13\%$ ($2\%$) from initially
$\sim 320$ for the entire AGB evolution of their solar 3\msun\ 
(4\msun) case and almost the same values for the \ose/\oac\ ratio
starting from initially $\sim 600$.  The evolutionary effect in the
overshoot model is significantly larger because the abundance of \ose\ 
in the dredged-up material is five to ten times larger if overshoot is
considered compared to models without overshoot. Over the first dozen
TPs available in \tab{3Meta1f} and \ref{4Meta1f}, the oxygen isotopic
ratios increase by about $44\%$ ($28\%$) for the 3\msun\ (4\msun)
case.  Similar to what has been said previously a comparison of
absolute numbers observed to the model predictions is misleading at
this stage. The initial model \ose/\osi\ ratios are probably too
small, possibly due to a wrong $\osi(p,\alpha)$ reaction rate. Also,
most of the observed data will belong to less massive stars for which
comparable model data do not yet exist. Clearly, this issue needs a
more detailed investigation. However, it seems that overshoot can
contribute to the solution of this problem.

\paragraph{Magnesium isotopic ratios}

The evolution of the magnesium isotopic ratios \mgvi/\mgfu\ and
\mgvi/\mgse\ are not only tracers of the TDUP but also of the
temperature in the He-burning shell during the thermal pulses. While
\mgvi\ is not produced or destructed in AGB stars the other Mg
isotopes are produced during the TP by $\alpha$ captures of \nezw. The
latter is abundant in the intershell as a result of two $\alpha$
captures on \nvi. Since the temperature at the bottom of the He-flash
convection zone is larger if overshoot is present
(\abb{fig:TP5-STRUC-all}) the reactions $\nezw(\alpha,n)\mgfu$ and
$\nezw(\alpha,\gamma)\mgse$ are more efficient. The 3\msun\ (4\msun)
sequence accordingly shows a decrease of \mgvi/\mgfu\ from $8.0$
($8.1$) at the first TP to $4.9$ ($3.9$) at the last TP computed.  The
\mgvi/\mgse\ ratio is $7.1$ ($7.0$) for the 3\msun\ (4\msun) sequence
at the first TP and decreases to $5.8$ ($5.3$) at the last computed TP
(see previous tables). Again, this decrease for the first dozen TP is
stronger than that predicted by Forestini \& Charbonnel
\cite*{forestini:97} for the entire evolution.

\section{Conclusions}
\label{sec:conclusions}

We have studied how the evolution of AGB stars is affected by
exponential diffusive overshoot at all convective boundaries. The main
differences between models with overshoot compared to models without
overshoot are
\begin{enumerate}
\item
  \begin{itemize}
  \item a depletion of helium and enhancement of carbon and oxygen in
    the intershell abundance distribution which is caused by,
  \item a deeper penetration of the He-flash convection zone into the
    C/O core below (intershell or fourth dredge-up),
\end{itemize}
\item
  \begin{itemize}
  \item larger energy generation by He-burning during the thermal
    pulse and consequently,
  \item higher temperature at the bottom of the He-flash convection
    zone and lower temperature at the top,
\end{itemize}
\item
  \begin{itemize}
  \item substantially increased dredge-up of material which has a
    modified abundance distribution compared to models without
    overshoot and as a result,
  \item a different evolution of the surface abundances showing a
    stronger signature of nucleosynthesis and
  \item a constant or decreasing core mass during some more advanced
    TPs,
\end{itemize}
\item
  \begin{itemize}
  \item a faster recovery of the hydrogen-burning shell after the TP
    and correlated with this,
  \item a faster recovery of the surface abundance after the TP
    compared to models without overshoot, and
\end{itemize}
\item the formation of a \cdr\ and a \nvi\ pocket at the envelope-core
  interface at the end of the dredge-up.
\end{enumerate}

These properties of AGB evolution models improve the understanding of
observationally obtained information about AGB stars. To date no
obvious contradictions of observational properties and AGB models with
overshoot have appeared. On the other hand, overshoot in AGB stellar
models (in particular that at the base of the He-flash convection
zone) has been an indispensable precondition for the starting point of
H-deficient post-AGB model sequences which can explain the observed
abundance ratios of these stars, in particular their oxygen abundance
\cite{herwig:99c}.  Therefore, overshoot should be considered an
important ingredient for the modeling of AGB stars.

\begin{acknowledgements}
  This work has been supported by the \emph{Deut\-sche
    For\-schungs\-ge\-mein\-schaft, DFG\/} (grants Scho\,394/13 and
  La\,587/16). I greatly appreciate previous improvements to the
  evolution code and stimulating support by T.\ Bl\"ocker. Further
  more I would like to thank R.\ Gallino, L.\ Koesterke, N.\ Langer
  and D.~Sch\"onberner for advice, helpful discussions and proof
  reading. The referee, A.\ Bressan, is acknowledged for his very
  useful comments.
\end{acknowledgements}


\begin{thebibliography}{}
  
\bibitem[\protect\astroncite{Alongi et~al.}{1991}]{alongi:91} Alongi,
  M., Bertelli, G., Bressan, A., and Chiosi, C., 1991, \newblock
  {A\&A} {244}, 95
  
\bibitem[\protect\astroncite{Alongi et~al.}{1993}]{alongi:93} Alongi,
  M., Bertelli, G., Bressan, A., Chiosi, C., Fagotto, F., Greggio, L.,
  and Nasi, E., 1993, \newblock {A\&A} {97}, 851
  
\bibitem[\protect\astroncite{Anders and Grevesse}{1989}]{anders:89}
  Anders, E. and Grevesse, N., 1989, \newblock {Geochim.\ Cosmochim.\ 
    Acta} {53}, 197
  
\bibitem[\protect\astroncite{Andersen et~al.}{1990}]{andersen:90}
  Andersen, J., Nordstr\"om, B., and Clausen, J.~V., 1990, \newblock
  {ApJ} {363}, L33
  
\bibitem[\protect\astroncite{{Aparicio} et~al.}{1991}]{aparicio:91}
  {Aparicio}, A., {Bertelli}, G., {Chiosi}, C., and {Garcia-Pelayo},
  J.~M., 1991, \newblock {A\&AS} {88}, 155
  
\bibitem[\protect\astroncite{Bl\"ocker}{1995}]{bloecker:95a}
  Bl\"ocker, T., 1995, \newblock {A\&A} {297}, 727
  
\bibitem[\protect\astroncite{Bl\"ocker et~al.}{2000}]{bloecker:99c}
  Bl\"ocker, T., Herwig, F., and Driebe, T., 2000, \newblock in F.
  D'Antona and R. Gallino (eds.), {The changes in abundances in AGB
    stars}, Mem. Soc. Astron. Ital., \newblock in press,
  astro-ph/0002455
  
\bibitem[\protect\astroncite{Bl\"ocker et~al.}{1997}]{bloecker:96b}
  Bl\"ocker, T., Herwig, F., Sch\"onberner, D., and Eid, M.~E., 1997,
  \newblock in {The Carbon Star Phenomenon}, IAU Symp.\ 177, Antalya,
  Turkey, \newblock in press
  
\bibitem[\protect\astroncite{B\"ohm-Vitense}{1958}]{boehm-vitense:58}
  B\"ohm-Vitense, E., 1958, \newblock {Z.\ Astrophys.} {46}, 108
  
\bibitem[\protect\astroncite{Boothroyd and
    Sackmann}{1988a}]{boothroyd:88c} Boothroyd, A.~I. and Sackmann,
  I.-J., 1988a, \newblock {ApJ} {328}, 653
  
\bibitem[\protect\astroncite{Boothroyd and
    Sackmann}{1988b}]{boothroyd:88} Boothroyd, A.~I. and Sackmann,
  I.-J., 1988b, \newblock {ApJ} {328}, 671
  
\bibitem[\protect\astroncite{Bressan et~al.}{1993}]{bressan:93}
  Bressan, A., Fagotto, F., Bertelli, G., and Chiosi, C., 1993,
  \newblock {A\&AS} {100}, 647
  
\bibitem[\protect\astroncite{Bressan et~al.}{1981}]{bressan:81}
  Bressan, A.~G., Bertelli, G., and Chiosi, C., 1981, \newblock {A\&A}
  {102}, 25
  
\bibitem[\protect\astroncite{Canuto}{1998}]{canuto:98a} Canuto, V.~M.,
  1998, \newblock {ApJ} {508}, L103
  
\bibitem[\protect\astroncite{Caughlan and Fowler}{1988}]{caughlan:88}
  Caughlan, G.~R. and Fowler, W.~A., 1988, \newblock {Atom.\ Data
    Nucl.\ Data Tables} {40}, 283, \newblock CF88
  
\bibitem[\protect\astroncite{D'Antona and
    Mazzitelli}{1996}]{d'antona:96} D'Antona, F. and Mazzitelli, I.,
  1996, \newblock {ApJ} {470}, 1093
  
\bibitem[\protect\astroncite{{De Marco} et~al.}{1998}]{demarco:97} {De
    Marco}, O., {Storey}, P.~J., and {Barlow}, M.~J., 1998, \newblock
  {MNRAS} {297}, 999
  
\bibitem[\protect\astroncite{Deng et~al.}{1996a}]{deng:96a} Deng, L.,
  Bressan, A., and Chiosi, C., 1996a, \newblock {A\&A} {313}, 145
  
\bibitem[\protect\astroncite{Deng et~al.}{1996b}]{deng:96b} Deng, L.,
  Bressan, A., and Chiosi, C., 1996b, \newblock {A\&A} {313}, 159
  
\bibitem[\protect\astroncite{Dreizler and Heber}{1998}]{dreizler:98}
  Dreizler, S. and Heber, U., 1998, \newblock {A\&A} {334}, 618
  
\bibitem[\protect\astroncite{El~Eid}{1994}]{eleid:94} El~Eid, M.,
  1994, \newblock {A\&A} {285}, 915
  
\bibitem[\protect\astroncite{Forestini and
    Charbonnel}{1997}]{forestini:97} Forestini, M. and Charbonnel, C.,
  1997, \newblock {A\&AS} {123}, 241
  
\bibitem[\protect\astroncite{Freytag et~al.}{1996}]{freytag:96}
  Freytag, B., Ludwig, H.-G., and Steffen, M., 1996, \newblock {A\&A}
  {313}, 497
  
\bibitem[\protect\astroncite{Frogel et~al.}{1990}]{frogel:90} Frogel,
  J.~A., Mould, J., and Blanco, V.~M., 1990, \newblock {ApJ} {352}, 96
  
\bibitem[\protect\astroncite{Frost and Lattanzio}{1996}]{frost:96}
  Frost, C.~A. and Lattanzio, J.~C., 1996, \newblock {ApJ} {473}, 383
  
\bibitem[\protect\astroncite{Gallino et~al.}{1998}]{gallino:97b}
  Gallino, R., Arlandini, C., Busso, M., Lugaro, M., Travaglio, C.,
  Straniero, O., Chieffi, A., and Limongi, M., 1998, \newblock {ApJ}
  {497}, 388
  
\bibitem[\protect\astroncite{Gallino et~al.}{1997}]{gallino:97a}
  Gallino, R., Busso, M., and Lugaro, M., 1997, \newblock in T.
  Bernatowitz and E. Zinner (eds.), {Astrophysical Implications of the
    Laboratory Study of Presolar Materials}, p. 115, AIP Conf.\ Ser.
  
\bibitem[\protect\astroncite{Gilroy}{1989}]{gilroy:89} Gilroy, K.~K.,
  1989, \newblock {ApJ} {347}, 835
  
\bibitem[\protect\astroncite{Habing}{1996}]{habing:96} Habing, H.~J.,
  1996, \newblock {A\&AR} {7}, 97
  
\bibitem[\protect\astroncite{Hamann}{1997}]{hamann:97} Hamann, W.-R.,
  1997, \newblock in H. Habing and H. Lamers (eds.), {Planetary
    Nebulae}, Vol. IAU Symp.\,180, p.~91, Kluwer
  
\bibitem[\protect\astroncite{{Harris} et~al.}{1987}]{harris:87}
  {Harris}, M.~J., {Lambert}, D.~L., {Hinkle}, K.~H., {Gustafsson},
  B., and {Eriksson}, K., 1987, \newblock {ApJ} {316}, 294
  
\bibitem[\protect\astroncite{{Harris} et~al.}{1985}]{harris:85}
  {Harris}, M.~J., {Lambert}, D.~L., and {Smith}, V.~V., 1985,
  \newblock {ApJ} {299}, 375
  
\bibitem[\protect\astroncite{Herwig and Bl\"ocker}{2000}]{herwig:99d}
  Herwig, F. and Bl\"ocker, T., 2000, \newblock in {The galactic Halo:
    From Globular Clusters to Field Stars}, Proc.  of the 35nd Li\`ege
  Int. Astr. Coll., Universtit\'e de Li\`ege, \newblock in press
  
\bibitem[\protect\astroncite{Herwig et~al.}{2000}]{herwig:00c} Herwig,
  F., Bl\"ocker, T., and Driebe, T., 2000, \newblock in F. D'Antona
  and R. Gallino (eds.), {The changes in abundances in AGB stars},
  Mem. Soc. Astron. Ital., \newblock in press, astro-ph/9912350
  
\bibitem[\protect\astroncite{Herwig et~al.}{1999a}]{herwig:99c}
  Herwig, F., Bl\"ocker, T., Langer, N., and Driebe, T., 1999a,
  \newblock {A\&A} {349}, L5, \newblock HBLD99
  
\bibitem[\protect\astroncite{Herwig et~al.}{1999b}]{herwig:98c}
  Herwig, F., Bl\"ocker, T., and Sch\"onberner, D., 1999b, \newblock
  in T.~L. Bertre, A. Lebre, and C. Waelkens (eds.), {AGB Stars}, IAU
  Symp.\,191, p.~41
  
\bibitem[\protect\astroncite{Herwig et~al.}{1997}]{herwig:97} Herwig,
  F., Bl\"ocker, T., Sch\"onberner, D., and {El Eid}, M.~F., 1997,
  \newblock {A\&A} {324}, L81
  
\bibitem[\protect\astroncite{Herwig et~al.}{1998}]{herwig:98b} Herwig,
  F., Sch\"onberner, D., and Bl\"ocker, T., 1998, \newblock {A\&A}
  {340}, L43
  
\bibitem[\protect\astroncite{Hollowell and Iben}{1988}]{hollowell:88}
  Hollowell, D. and Iben, Jr., I., 1988, \newblock {ApJ} {333}, L25
  
\bibitem[\protect\astroncite{Hurlburt et~al.}{1986}]{hurlburt:86}
  Hurlburt, N.~E., Toomre, J., and Massaguer, J.~M., 1986, \newblock
  {ApJ} {311}, 563
  
\bibitem[\protect\astroncite{Hurlburt et~al.}{1994}]{hurlburt:94}
  Hurlburt, N.~E., Toomre, J., Massaguer, J.~M., and Zahn, J.-P.,
  1994, \newblock {ApJ} {421}, 245
  
\bibitem[\protect\astroncite{Iben}{1975}]{iben:75} Iben, Jr., I.,
  1975, \newblock {ApJ} {196}, 525
  
\bibitem[\protect\astroncite{Iben}{1976}]{iben:76} Iben, Jr., I.,
  1976, \newblock {ApJ} {208}, 165
  
\bibitem[\protect\astroncite{Iben}{1977}]{iben:77} Iben, Jr., I.,
  1977, \newblock {ApJ} {217}, 788
  
\bibitem[\protect\astroncite{Iben}{1999}]{iben:99} Iben, Jr., I.,
  1999, \newblock in T.~L. Bertre, A. Lebre, and C. Waelkens (eds.),
  {AGB Stars}, IAU Symp.\,191, p. 591
  
\bibitem[\protect\astroncite{Iben and Renzini}{1983}]{iben:83b} Iben,
  Jr., I. and Renzini, A., 1983, \newblock {ARA\&A} {21}, 271
  
\bibitem[\protect\astroncite{Iben and Truran}{1978}]{iben:78} Iben,
  Jr., I. and Truran, J.~W., 1978, \newblock {ApJ} {220}, 980
  
\bibitem[\protect\astroncite{Iben et~al.}{1996}]{iben:96} Iben, Jr.,
  I., Tutukov, A.~V., and Yungelson, L.~R., 1996, \newblock {ApJ}
  {456}, 750
  
\bibitem[\protect\astroncite{Iglesias and Rogers}{1996}]{iglesias:96}
  Iglesias, C.~A. and Rogers, F.~J., 1996, \newblock {ApJ} {464}, 943
  
\bibitem[\protect\astroncite{Kippenhahn and
    Weigert}{1990}]{kippenhahn:90} Kippenhahn, R. and Weigert, A.,
  1990, \newblock {Stellar structure and evolution}, \newblock
  Springer, Berlin
  
\bibitem[\protect\astroncite{{Koesterke} and
    {Hamann}}{1997}]{koesterke:97b} {Koesterke}, L. and {Hamann},
  W.~R., 1997, \newblock {A\&A} {320}, 91
  
\bibitem[\protect\astroncite{Kozhurina-Platais
    et~al.}{1997}]{kozhurina:97} Kozhurina-Platais, V., Demarque, P.,
  Platais, I., Orosz, J.~A., and Barnes, S., 1997, \newblock {AJ}
  {113(3)}, 1045
  
\bibitem[\protect\astroncite{Landr\'e et~al.}{1990}]{landre:90}
  Landr\'e, V., Prantzos, N., Aguer, P., Bogaert, G., Lefebvre, A.,
  and Thibaud, J., 1990, \newblock {A\&A} {240}, 85
  
\bibitem[\protect\astroncite{Langer}{1986}]{langer:86} Langer, N.,
  1986, \newblock {A\&A} {164}, 45
  
\bibitem[\protect\astroncite{Langer et~al.}{1985}]{langer:85} Langer,
  N., {El Eid}, M., and Fricke, K.~J., 1985, \newblock {A\&A} {145},
  179
  
\bibitem[\protect\astroncite{{Langer} et~al.}{1983}]{langer:83}
  {Langer}, N., {Fricke}, K.~J., and {Sugimoto}, D., 1983, \newblock
  {A\&A} {126}, 207+
  
\bibitem[\protect\astroncite{Langer et~al.}{1999}]{langer:99} Langer,
  N., Heger, A., Wellstein, S., and Herwig, F., 1999, \newblock {A\&A}
  {346}, L37
  
\bibitem[\protect\astroncite{Lattanzio}{1986}]{lattanzio:86}
  Lattanzio, J.~C., 1986, \newblock {ApJ} {311}, 708
  
\bibitem[\protect\astroncite{Lattanzio}{1989}]{lattanzio:89a}
  Lattanzio, J.~C., 1989, \newblock {ApJ} {344}, L25
  
\bibitem[\protect\astroncite{Lattanzio and
    Boothroyd}{1997}]{lattanzio:97} Lattanzio, J.~C. and Boothroyd,
  A.~I., 1997, \newblock in T. Bernatowitz and E. Zinner (eds.),
  {Astrophysical Implications of the Laboratory Study of Presolar
    Materials}, p.~85, AIP Conf.\ Ser.
  
\bibitem[\protect\astroncite{Maeder}{1975}]{maeder:75} Maeder, A.,
  1975, \newblock {A\&A} {40}, 303
  
\bibitem[\protect\astroncite{Maeder and Meynet}{1989}]{maeder:89}
  Maeder, A. and Meynet, G., 1989, \newblock {A\&A} {210}, 155
  
\bibitem[\protect\astroncite{Marigo et~al.}{1996}]{marigo:96} Marigo,
  P., Bressan, A., and Chiosi, C., 1996, \newblock {A\&A} {313}, 545
  
\bibitem[\protect\astroncite{Marigo et~al.}{1999}]{marigo:99a} Marigo,
  P., Girardi, L., Weiss, A., and Groenewegen, M. A.~T., 1999,
  \newblock {A\&A} {351}, 161, \newblock in press
  
\bibitem[\protect\astroncite{Mazzitelli et~al.}{1999}]{mazzitelli:99}
  Mazzitelli, I., D'Antona, F., and Ventura, P., 1999, \newblock
  {A\&A} {348}, 846
  
\bibitem[\protect\astroncite{Mermilliod and
    Maeder}{1986}]{mermilliod:86} Mermilliod, J.~C. and Maeder, A.,
  1986, \newblock {A\&A} {158}, 45
  
\bibitem[\protect\astroncite{Mowlavi}{1999}]{mowlavi:99} Mowlavi, N.,
  1999, \newblock {A\&A} {344}, 617
  
\bibitem[\protect\astroncite{Napiwotzki et~al.}{1991}]{napiwotzki:91}
  Napiwotzki, R., Sch\"onberner, D., and Weidemann, V., 1991,
  \newblock {A\&A} {243}, L5
  
\bibitem[\protect\astroncite{Paczy\'nski}{1977}]{paczynski:77}
  Paczy\'nski, B., 1977, \newblock {ApJ} {214}, 812
  
\bibitem[\protect\astroncite{Pinsonneault}{1997}]{pinsonneault:97}
  Pinsonneault, M., 1997, \newblock {ARA\&A} {35}, 557
  
\bibitem[\protect\astroncite{Refsdal and Weigert}{1970}]{refsdal:70}
  Refsdal, S. and Weigert, A., 1970, \newblock {Z.\ Astrophys.} {6},
  426
  
\bibitem[\protect\astroncite{Renzini}{1987}]{renzini:87} Renzini, A.,
  1987, \newblock {A\&A} {188}, 49
  
\bibitem[\protect\astroncite{Roxburgh}{1978}]{roxburgh:78} Roxburgh,
  I.~W., 1978, \newblock {A\&A} {65}, 281
  
\bibitem[\protect\astroncite{{Sackmann}}{1980}]{sackmann:80}
  {Sackmann}, I.~J., 1980, \newblock {ApJ Lett.} {241}, L37
  
\bibitem[\protect\astroncite{{Salasnich} et~al.}{1999}]{salasnich:99}
  {Salasnich}, B., {Bressan}, A., and {Chiosi}, C., 1999, \newblock
  {A\&A} {342}, 131
  
\bibitem[\protect\astroncite{Schaller et~al.}{1992}]{schaller:92}
  Schaller, G., Schaerer, D., Meynet, G., and Maeder, A., 1992,
  \newblock {A\&AS} {96}, 269
  
\bibitem[\protect\astroncite{Sch\"onberner}{1979}]{schoenberner:79}
  Sch\"onberner, D., 1979, \newblock {A\&A} {79}, 108
  
\bibitem[\protect\astroncite{Sch\"onberner}{1996}]{schoenberner:96}
  Sch\"onberner, D., 1996, \newblock in C.~S. Jeffery and U. Heber
  (eds.), {Hydrogen-Deficient Stars}, Vol.~96 of {ASP Conf. Series},
  p. 433
  
\bibitem[\protect\astroncite{Schr\"oder et~al.}{1997}]{schroeder:97}
  Schr\"oder, K.-P., Pols, O.~R., and Eggleton, P., 1997, \newblock
  {MNRAS} {285}, 696
  
\bibitem[\protect\astroncite{Schwarzschild and
    H\"arm}{1965}]{schwarzschild:65} Schwarzschild, M. and H\"arm, R.,
  1965, \newblock {ApJ} {142}, 855
  
\bibitem[\protect\astroncite{Shaviv and Salpeter}{1973}]{shaviv:73}
  Shaviv, G. and Salpeter, E., 1973, \newblock {ApJ} {184}, 191
  
\bibitem[\protect\astroncite{Smith and Lambert}{1990}]{smith:90}
  Smith, V.~V. and Lambert, D.~L., 1990, \newblock {ApJS} {72}, 387
  
\bibitem[\protect\astroncite{Smith et~al.}{1987}]{smith:87} Smith,
  V.~V., Lambert, D.~L., and McWilliam, A., 1987, \newblock {ApJ}
  {320}, 862
  
\bibitem[\protect\astroncite{Stein and Nordlund}{1998}]{stein:98}
  Stein, R.~F. and Nordlund, A., 1998, \newblock {ApJ} {499}, 914
  
\bibitem[\protect\astroncite{Stothers}{1991}]{stothers:91} Stothers,
  R., 1991, \newblock {ApJ} {383}, 820
  
\bibitem[\protect\astroncite{Straniero et~al.}{1997}]{straniero:97}
  Straniero, O., Chieffi, A., Limongi, M., Busso, M., Gallino, R., and
  Arlandini, C., 1997, \newblock {ApJ} {478}, 332
  
\bibitem[\protect\astroncite{Straniero et~al.}{1995}]{straniero:95}
  Straniero, O., Gallino, R., Busso, M., Chieffi, A., Raiteri, C.~M.,
  Salaris, M., and Limongi, M., 1995, \newblock {ApJ} {440}, L85
  
\bibitem[\protect\astroncite{Vassiliadis and
    Wood}{1993}]{vassiliades:93} Vassiliadis, E. and Wood, P., 1993,
  \newblock {ApJ} {413}, 641
  
\bibitem[\protect\astroncite{Ventura et~al.}{1998}]{ventura:98}
  Ventura, P., Zeppieri, A., Mazzitelli, I., and {D'Antona}, F., 1998,
  \newblock {A\&A} {334}, 953
  
\bibitem[\protect\astroncite{Wagenhuber and
    Groenewegen}{1998}]{wagenhuber:98} Wagenhuber, J. and Groenewegen,
  M. A.~T., 1998, \newblock {A\&A} {340}, 183
  
\bibitem[\protect\astroncite{Wallerstein
    et~al.}{1997}]{wallerstein:97} Wallerstein, G., Iben, Jr., I.,
  Parker, P., Boesgard, A.~M., Hale, G.~M., Champagne, A.~E., Barnes,
  C.~A., K\"appeler, F., Smith, V.~V., Hoffmann, R.~D., Timmes, F.~X.,
  Sneden, C., Boyd, R.~N., Meyer, B.~S., and Lambert, D.~L., 1997,
  \newblock {Rev.\ Mod.\ Phys.} {69(4)}, 995
  
\bibitem[\protect\astroncite{Wallerstein and
    Knapp}{1998}]{wallerstein:98} Wallerstein, G. and Knapp, G.~R.,
  1998, \newblock {ARA\&A} {36}, 369
  
\bibitem[\protect\astroncite{Weaver and Woosley}{1993}]{weaver:93}
  Weaver, T.~A. and Woosley, S.~E., 1993, \newblock {Physics Reports}
  {40(227)}, 65
  
\bibitem[\protect\astroncite{Weidemann}{2000}]{weidemann:00}
  Weidemann, V., 2000, \newblock {A\&A}, \newblock submitted
  
\bibitem[\protect\astroncite{Weigert}{1966}]{weigert:66} Weigert, A.,
  1966, \newblock {Z.\ Astrophys.} {64}, 395
  
\bibitem[\protect\astroncite{Werner et~al.}{1999}]{werner:98} Werner,
  K., Dreizler, S., Rauch, T., Koesterke, L., and Heber, U., 1999,
  \newblock in T.~L. Bertre, A. Lebre, and C. Waelkens (eds.), {AGB
    Stars}, IAU Symp.\,191, p. 493
  
\bibitem[\protect\astroncite{{Wood}}{1981}]{wood:81} {Wood}, P.~R.,
  1981, \newblock in I. Iben, Jr. and A. Renzini (eds.), {Physical
    Processes in Red Giants}, p. 135
  
\bibitem[\protect\astroncite{Xiong}{1985}]{xiong:85} Xiong, D.~R.,
  1985, \newblock {A\&A} {150}, 133
  
\bibitem[\protect\astroncite{{Zuckerman} and
    {Aller}}{1986}]{zuckermann:86} {Zuckerman}, B. and {Aller}, L.~H.,
  1986, \newblock {ApJ} {301}, 772

\end{thebibliography}
\end{document}